\documentclass[journal,twoside,web]{ieeecolor}
\usepackage{generic} 
\usepackage{amsmath,graphicx}
\usepackage{adjustbox}
\usepackage[para,online,flushleft]{threeparttable}
\usepackage{svg}
\usepackage{amsmath,amssymb,amsfonts}
\let\labelindent\relax
\usepackage{enumitem}
\usepackage{cite}
\usepackage{float}

\usepackage{amsmath}

\usepackage {cancel} 

\usepackage{xcolor}
\usepackage{hyperref}
\hypersetup{hidelinks=true}
\usepackage[utf8]{inputenc}
\usepackage{pgfplots}
\usepackage{caption}
\usepackage{subcaption}
\usepackage{pifont}
\usepackage{tikz}
\usepackage[utf8]{inputenc}
\usepackage{pgfplots}
\DeclareUnicodeCharacter{2212}{−}
\usepgfplotslibrary{groupplots,dateplot}
\usetikzlibrary{patterns,shapes.arrows}
\pgfplotsset{compat=newest}

\usepackage{multirow,bigdelim,dcolumn,booktabs}
\newcommand{\cmark}{\ding{51}}%
\newcommand{\xmark}{\ding{55}}%
\DeclareUnicodeCharacter{2212}{−}
\usepgfplotslibrary{groupplots,dateplot}
\usetikzlibrary{patterns,shapes.arrows}
\pgfplotsset{compat=newest}

\definecolor{brown1926061}{RGB}{192,60,61}
\definecolor{darkgray176}{RGB}{176,176,176}
\definecolor{darkslategray61}{RGB}{61,61,61}
\definecolor{lightgray204}{RGB}{204,204,204}
\definecolor{peru22412844}{RGB}{224,128,44}
\definecolor{seagreen5814558}{RGB}{58,145,58}
\definecolor{sg}{RGB}{58,145,58}
\definecolor{sr}{RGB}{255,102,102}
\definecolor{steelblue49115161}{RGB}{49,115,161}
 
\makeatletter
\newcommand{\startappendix}{%
  \cleardoublepage
  \appendix
  \phantomsection\addcontentsline{toc}{part}{APPENDIX}
  \part*{APPENDIX}
  \let\@makechapterhead\@makeschapterhead
}
\makeatother

\def\BibTeX{{\rm B\kern-.05em{\sc i\kern-.025em b}\kern-.08em
    T\kern-.1667em\lower.7ex\hbox{E}\kern-.125emX}}
\title{A Strong and Simple Deep Learning Baseline for BCI Motor Imagery decoding}
\author{Yassine El Ouahidi,~\IEEEmembership{Student,~IEEE,} Vincent Gripon,~\IEEEmembership{Senior,~IEEE,} Bastien Pasdeloup, Ghaith Bouallegue, Nicolas Farrugia and Giulia Lioi
\thanks{Thanks to the Brittany region, as well as the ANR program AI@IMT for their support.}
\thanks{Authors are with the Mathematical and Electrical Engineering
Department, IMT Atlantique, Lab-STICC, UMR CNRS 6285, F-29238 Brest, France (e-mail: name.surname@imt-atlantique.fr).}
}

\newcommand*{\refname}{Bibliography}

\begin{document}
%
\maketitle
\begin{abstract}

We propose EEG-SimpleConv, a straightforward 1D convolutional neural network for Motor Imagery decoding in BCI. Our main motivation is to propose a simple and performing baseline to compare to, using only very standard ingredients from the literature. We evaluate its performance on four EEG Motor Imagery datasets, including simulated online setups, and compare it to recent Deep Learning and Machine Learning approaches. EEG-SimpleConv is at least as good or far more efficient than other approaches, showing strong knowledge-transfer capabilities across subjects, at the cost of a low inference time. We advocate that using off-the-shelf ingredients rather than coming with ad-hoc solutions can significantly help the adoption of Deep Learning approaches for BCI. We make the code of the models and the experiments accessible.

\end{abstract}


\section{Introduction}

Brain-computer interfaces (BCI) have gained significant attention in research applications, enabling direct interaction between individuals and their environment based on brain signals. Among BCI paradigms, Motor Imagery (MI) holds particular significance for its potential social and medical impacts (\emph{e.g.}, through assistive technologies for individuals with impaired motor skills, independent communication or post-stroke motor rehabilitation~\cite{nam2018brain,lioi2020multi}).

In recent years, the integration of Machine Learning has made a significant impact on BCI, as Machine Learning models possess the ability to adapt to specific tasks. This interdisciplinary field has witnessed substantial progress and a plethora of algorithms specifically designed for EEG based BCIs have been proposed, with Riemannian geometry and Common Spatial Pattern (CSP)~\cite{koles1990spatial} based methods with automatic features selection the most common and effective approaches (see~\cite{lotte2018review} for an extensive review).

However, current BCI systems suffer from several limitations as they have poor reliability, require a long calibration for each subject (or each session), and in some cases extensive training. These limitations arise from the noisy and non-stationary nature of EEG signals, and their unique characteristics for each individual, resulting in significant inter-subject variations in brain activity patterns. This variability poses a challenge in developing BCI systems that generalize to new users, limiting BCI usability outside research laboratories.

Methods based on Deep Learning, with their generalization and transfer capabilities, hold great potential for achieving systems that can generalize across several subjects by exploiting information from multi-subjects datasets.
Nevertheless, the utilization of Deep Learning in BCIs has not yet demonstrated convincing performance and robustness, and the literature continues to encounter challenges~\cite{lotte2018review}. One key concern is the biased evaluation of Deep Learning models performance, often trained on one dataset and assessed on single subjects or datasets, which restricts the demonstration of their ability to generalize to diverse subjects or EEG recording setups. Furthermore, the lack of clarity regarding the rationale behind specific architectural and optimization choices, makes it difficult to rigorously evaluate the added value of single ingredients. Finally, the practice of sharing the models implementation to allow for benchmarking is not ubiquitous in the BCI community, adding to the existing challenges. 

In this work we address these limitations and propose EEG-SimpleConv, a 1D convolutional neural network with a simple architecture made only of off-the-shelf standard ingredients. Our network is designed specifically for EEG-based MI tasks. To assess its performance, we conduct evaluations on four EEG MI datasets and compare it to state-of-the-art Deep Learning models commonly used in this domain. In addition, we justify and validate with ablation studies the key components of the architecture and its training routine.

The main contributions of this paper are:
\begin{itemize}
    \item A novel and straightforward 1D convolutional architecture for MI decoding, achieving state-of-the-art performance on widely recognized open source MI datasets. 
    \item An ablation study showcasing the merit of all included ingredients.    
    \item The code to reproduce all experiments in this paper, available here: \url{https://github.com/elouayas/EEGSimpleconv}.
\end{itemize}

\section{Related work}

Deep Neural Networks have introduced a promising way of exploiting and transferring knowledge across sessions and subjects, and have emerged as a viable candidate solution to address the limitations of conventional Machine Learning methods. Recent works have investigated their potential for different BCI paradigms and numerous architectural designs have been proposed. In this section we will discuss various contributions using Deep Learning approaches for BCI.

\subsection{Convolutionals Neural Networks (CNN) architectures}
CNNs present several notable benefits, encompassing end-to-end learning, effective pattern analysis, and customizable depth and width. Notably, in diverse domains, CNNs have demonstrated superior performance compared to previous methodologies, particularly in tasks associated with vision. Due to their effectiveness in capturing both spatial and time-frequency characteristics of neural signals, CNNs have become the predominant architecture for decoding MI signals.

\subsubsection{1D CNN}

The most straightforward and intuitive approach for leveraging convolution in the analysis of EEG signals is through the utilization of 1D convolution. By employing temporal convolutions, it becomes feasible to capture temporal patterns within the data. Other types of 1D convolution have also been proposed such as 1D spatial convolutions~\cite{li2019channel}, dilated 1D convolutions and Mixed-Scale-Conv blocks~\cite{lun2020simplified}. Other works introduced more complex layers such as Residual blocks~\cite{he2016deep}, Inception blocks~\cite{szegedy2015going}, and Squeeze and Excitation blocks~\cite{hu2018squeeze} forming a Multi-branch Multi-scale 1D CNN in~\cite{liu2023compact}.
In our work we only rely on 1D-temporal convolutions in a simple architecture, resembling early vision CNNs. A similar model has been proposed in~\cite{mattioli20221d} for BCI.

\subsubsection{2D CNN}

While 1D convolutions provide an intuitive approach for capturing temporal patterns, early architectures for BCI primarily used 2D convolutions, treating the spatio-temporal input signal as an image. DeepConvNet~\cite{schirrmeister2017deep}, originally designed in~\cite{cecotti2010convolutional} for P300 tasks, was experimented on MI tasks as well. This architecture consists of two convolutional layers for extracting temporal and spatial features. A variant proposed by the same authors, Shallow ConvNet~\cite{schirrmeister2017deep}, demonstrated superior performance compared to the FBCSP algorithm~\cite{ang2008filter}. Both works highlight the separation of temporal and spatial convolutions as a central feature of the architecture. Subsequently EEGNet, an evolution of these models, was proposed~\cite{lawhern2018eegnet}, employing 2D convolutions to achieve a more compact, faster, and efficient architecture. EEGNet incorporates additional components such as dropout layers, batch normalisation, kernel constraints, and max-pooling strides. These three architectures, with EEGNet being the standard, remain the most popular models for MI classification. Very similar architectures have also been proposed in~\cite{dose2018end,tibrewal2022classification}.

Numerous variants of these three architectures have then been proposed, incorporating increasingly complex modules such as Residual blocks in Residual ConvNet\cite{schirrmeister2017deep}. Temporal Convolution (TC) modules (Residual blocks of causal convolutions with dilation) were proposed in EEG-TCNet~\cite{ingolfsson2020eeg}. A similar TC module with the addition of a dense spatial module (densely connected convolutions aiming at capturing spatial features) was introduced in TIDNet~\cite{kostas2020thinker}. An Inception module (parallel stream of separable convolutions to enhance spatial feature extraction) inspired by Inception-Time Neural Network~\cite{ismail2020inceptiontime}) was also proposed in EEGInception~\cite{zhang2021eeg} and EEG-ITNet~\cite{salami2022eeg} (in combination with TC modules). Inception and Xception modules~\cite{chollet2017xception} (parallel streams of separable convolutions) are found in MI-EEGNet~\cite{riyad2021mi}.

One of the primary motivation behind the incorporation of various modules in many of the above-mentioned architectures is reducing network size by minimizing the number of parameters. For instance, the use of dilated convolutions justifies the expansion of receptive fields without excessively deepening the architecture. This rationale stems from the requirement that MI decoding models should ideally be fast for real-life BCI applications. However, within the domain of Deep Learning, a low number of parameters does not necessarily guarantee low latency. We will demonstrate that, a straightforward architecture, despite having a higher number of parameters, can still exhibit remarkably low latency and remain highly competitive and practical for real-world applications.

\subsection{Other types of Neural Networks}

\subsubsection{Recurrent Neural Networks and Transformers}

Despite the rise of CNN-based architectures for MI-based BCI decoding, Recurrent Neural Networks (RNN) and Long Short-Term Memory (LSTM) architectures have been explored with limited success for BCI applications~\cite{luo2018exploring,wang2018lstm}. These recurrent architectures have evolved into transformers~\cite{vaswani2017attention}, and have also been explored for MI decoding in recent works~\cite{xie2022transformer,song2021transformer,song2022eeg}. 

\subsubsection{Riemannian geometry based Neural Networks}

In MI decoding, alternative architectures have been proposed. SPDNet~\cite{huang2017riemannian} exploits Riemannian geometry and takes as input Symmetric Positive Definite (SPD) EEG covariance matrices instead of raw EEG signals, and addresses the non-stationary and variant local statistics of EEG signals by preserving the SPD structure across layers. Variants of this architecture exploiting SPD manifolds using tensors (Tensor-SPDNet~\cite{ju2022tensor}) or Graph Neural Networks (GNN-SPDNet~\cite{ju2022graph}) have also been proposed. A couple of interesting recent works have combined the advantages of Deep Learning and Riemannian methods to improve generalization performances~\cite{NEURIPS2022_28ef7ee7,wilson2022deep}.~\cite{wilson2022deep} is a promising architecture (CNN-SPDNet)  that can be considered a two-step neural network rather than a traditional hybrid architecture due to its non-end-to-end training approach.

\subsubsection{More complex architectures}

Lastly, a variety of more intricate architectures have been proposed. Among them are hybrid architecture such as CNN-LSTM architecture~\cite{khademi2022transfer} and multi-branch architectures. The latter combine multiple neural networks in parallel, typically three or four, such as Multi-branch 3D CNN~\cite{zhao2019multi} where multiple 3D CNNs are combined together, CCNN~\cite{amin2019deep} with multiple 2D CNNs, or CMO-CNN~\cite{liu2023compact} with multiple 1D CNNs.
In multi-view architectures~\cite{zhang2023multi,li2023parallel,wu2019parallel} the idea is quite similar to multi-branch but each input is a different representation of the raw data (\emph{e.g.}, different time-frequency ranges). 

To summarize, a wide range of architectures have been proposed for BCI decoding ranging from simple 1D CNN to more complex architectures. However, the rationale underlying specific architectural choices is often unclear. The most common approach is to add modules on top of an architecture that works, with domain motivation. But it's often difficult to understand the impact and contribution of each of these modules to the classification performances. In this study we go back several steps in time, showing that state-of-the-art performances are achievable with a very simple architecture. We also insist on having rigorous and numerous evaluation steps in order to serve as a reliable benchmark for other works. 

Secondly, we show in Section~\ref{sec:time_inference}, that the proposed network, EEG-SimpleConv, based only on simple 1D temporal convolutions, is very fast, despite its size, possibly allowing to deploy Deep Learning methods for online BCI uses.

\section{contributions}

\begin{figure*}
\centering
\includegraphics[width=0.95\linewidth]{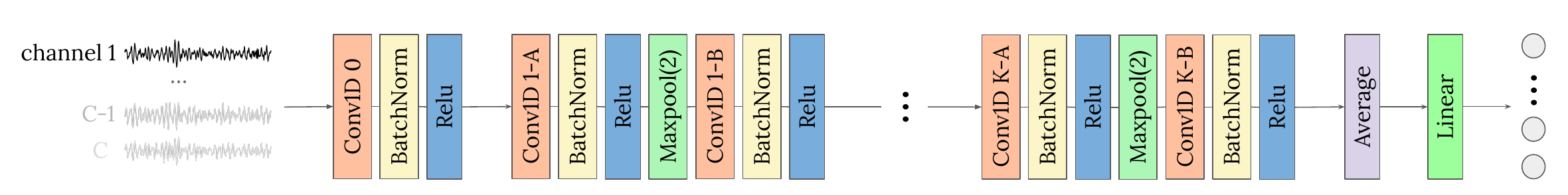}
\caption{EEG-SimpleConv architecture}
\label{fig:archi}
\end{figure*}

EEG-SimpleConv is largely inspired by classical vision CNNs.
Its architecture is illustrated in Fig.~\ref{fig:archi}. The main idea is to stack convolutional layers. At some steps, the processed signal is downsampled in time using max pooling, and the number of feature maps is increased to compensate for the reduction of dimension space. An input convolutional layer is used to embed the signal in a space with the appropriate number of feature maps (a hyperparameter of our architecture). At the end of the architecture, the signal is averaged over time, making the full pipeline architecture invariant to temporal shifts of its inputs, as well as allowing to process inputs with various lengths. A final fully connected layer maps the averaged features to a logit space, which dimension depends on the number of classes in the considered task.

The main difference with vision CNNs is that we use 1D convolutions instead of 2D ones. Input feature maps correspond to the various electrodes, embodying a spatial dimension of the signal. Yet, compared to many alternatives, we do not exploit any topology or connectivity about the electrodes.

Let us add a small remark on max pooling. In 2D ResNet architectures, max pooling reduces the spatial dimension of processed feature vectors by 4 (2x2), and is compensated by multiplying the number of feature maps by 2. This choice has the interest of maintaining the flops (floating point operations per second) constant throughout the architecture, as it evolves linearly with the number of spatial dimensions and quadratically with the number of feature maps. To mimic this behavior, we decided to increase the number of feature maps by $\sqrt{2}$ each time we reduce time resolution by 2 using max pooling.

By using a sequence of small 1D convolutional filters stacked on top of each other, the network becomes deeper and incorporates more non-linearities. This approach allows EEG-SimpleConv to capture intricate patterns and dependencies within the EEG signals, enhancing its ability to extract meaningful features for classification tasks. Stacking convolutions also has the advantage of increasing the receptive field, while maintaining the number of flops reasonable.

Similarly to vision architectures, we use batch normalisation layers after each convolution, and use relu activation functions.

Batch normalisation is applied after each convolutional layer to normalize activations within mini-batches, effectively reducing distribution shift and enhancing training stability. In our training routine, detailed in Section~\ref{sec:ingredient_routine}, we propose to use batch normalisation in a novel manner, allowing seamless adaptation to new users or sessions during model inference.

\subsection{EEG-SimpleConv architecture}
\label{sec:architecture}

Our architecture is fully determined by knowing:
\begin{itemize}
\item The width $W$, i.e. the number of feature maps at the output of the embedding convolutional layer -- all remaining number of feature maps is linearly depending on it;
\item The depth $2\times K+2$ of the architecture, that is the total number of layers, including the initial embedding convolution and the final fully connected layer;
\item The kernel size $S$ of convolutions;
\item The precise locations where to use max pooling and compensate by multiplying the number of feature maps by $\sqrt{2}$. In our experiments, we reduce the search space by always using max pooling every two convolutions (the embedding convolution being excluded).
\end{itemize}

Guidelines on how to choose the right values for thoses parameters are available in Appendix~\ref{app:tune_hyperparam}.

\subsection{Key ingredients of the training routine}
\label{sec:ingredient_routine}
In this section, we outline the components of our training routine. Prior to training, we adopt a minimalist preprocessing approach for our EEG data, aiming to provide the model with the most unaltered input data possible. Subsequently, we incorporate two types of ingredients into our training process, that are common practice in modern Deep Learning routines. The first type focuses on data normalisation at various levels, while the second type pertains to regularization techniques. 


 \subsubsection{EEG Signal Preprocessing}
Unlike traditional Machine Learning models, the advantage of Deep Learning lies in its ability to work with raw data and minimize the preprocessing steps, allowing the model to automatically learn relevant features. In line with recent results~\cite{Delorme2023}, we minimally preprocess raw EEG data by applying a simple high-pass filter at 0.5Hz.

In addition, we resample the signals to a lower frequency than acquisition sampling frequency to accelerate the training and inference processes by working with shorter signals. Resampling also indirectly acts as a low-pass filter by filtering out high-frequency components. The resampling frequency is selected such that the low-pass filter still include the relevant frequency bands for the classification of motor imagery signals, typically up to 40-50Hz.

By incorporating these preprocessing steps, we aim to optimize the training and inference efficiency of our neural networks while preserving the key frequency information necessary for accurate classification of motor imagery signals.

\subsubsection{Normalisation}

\begin{itemize}
    \item \textbf{Euclidean Alignment (EA)}: Machine Learning models perform best when training and testing sets belong to the same distribution. However, EEG data from different subjects and sessions have varying distributions, reducing BCI model performance.
     EA addresses this issue by projecting the data into a domain-invariant space based on the covariance matrix, ensuring consistency across subjects and sessions by making data more homogeneous and less scattered~\cite{he2019transfer}. Let's consider a subject with $n$ trials, where each trial, consists of $C \times T$ samples with $C$ the number of channels and $T$ the number of time samples, and is represented by matrix ${\bf X}_i \in \mathbb R^{c \times t}$, we have:

    \begin{equation}
    {\bf \bar{R}}=\frac{1}{n} \sum_{i=1}^{n} {\bf X}_i {\bf X}^{T}_i 
    \end{equation}
    Alignment is performed by using the matrix square root of the arithmetic mean:
    \begin{equation}
    {\bf \tilde{X}}_i= {\bf \bar{R}}^{-\frac{1}{2}} {\bf X}_i
    \end{equation}

    \item \textbf{Standardization}: it is the most common normalisation technique used while dealing with numerical data. It consists of transforming the data into a standard format with a mean of zero and a standard deviation of one, scaling it across subjects and sessions. Even though, EA has a comparable scaling effect, we still perform standardization, to make sure we have these properties when we ablate EA of our pipeline.
    
    \item \textbf{Session statistics} instead of Subject statistics: EEG signals recorded from the same individual in different recording sessions shows variations. These variations can arise due to a variety of factors, such as changes in physiological and psychological states, electrode placement, experimental conditions, and environmental factors. 
    To overcome theses variations, we perform the various normalisations proposed in this section using session statistics instead of subject statistics.

    \item \textbf{Batch Normalisation (BN) ``trick''} - Recompute Batch normalisation statistics on the tests sets: BN is a widely adopted and effective technique in Deep Learning. However, its behavior varies between training and testing phases. In training, running mean and variance are calculated and updated per batch. During evaluation, the learned and stored statistics from training are utilized to normalize the input. In our approach, we innovate the use of BN. During testing, we calculate statistics directly from the test sets rather than relying on stored training statistics. By processing all data from a single user (or session) as a batch, we obtain user-specific (or session-specific) statistics. This adaptation enhances the model's ability to adjust to individual users, as the BN layers incorporate personalized user-specific (or session-specific) statistics.
      
\end{itemize}
    It is worth mentioning that both EA and BN trick cannot be used as described in online scenarios.
    
\subsubsection{Regularization}
\begin{itemize}
    \item \textbf{Mixup}: it is an effective form of data augmentation technique in Deep Learning that addresses issues like overfitting, limited generalization, and vulnerability to adversarial attacks~\cite{zhang2017mixup}. It improves model performance by generating synthetic training examples through linear interpolation of both input samples ($\tilde{x}$) and corresponding labels ($\tilde{y}$). By incorporating these mixed examples in training, Mixup promotes robust and generalizable model representations. It acts as a regularization method, introducing diversity and implicit noise, preventing overfitting to specific examples.

    \item \textbf{Subject-wise Regularisation}: To tackle variations in EEG signals across subjects, we propose a novel subject-wise regularization. This involves adding a classification head to identify the subject alongside the existing task classification head. The additional head helps the model understand inter-subject variations. Using a standard cross-entropy loss for training, the model learns to accurately classify subject labels associated with EEG signals. This regularization improves the model's ability to handle subject variability.

\end{itemize}

\subsubsection{Training procedure details}
We train our neural network with the Adam optimizer, a popular choice for Deep Learning, for 50 epochs. Adam combines the advantages of both adaptive learning rates and momentum-based optimization techniques. At the 40th epoch, we apply a learning rate decay (by a factor 0.1) to enhance convergence. The standard cross-entropy loss function is used, suitable for multi-class classification, minimizing discrepancies between predicted probabilities and true labels for EEG signals. For comprehensive evaluation, we employ a rigorous cross-testing approach, dividing the dataset into folds. By iterating through each fold, we evaluate the model's performance accurately on the whole dataset. We repeat the training five times with different initial weights, averaging performance metrics for reliable estimates. As the datasets are balanced, accuracy is used for model evaluation. Detailed evaluation methodologies process can be found in Section~\ref{sec:evaluation}.

\section{Datasets and Evaluation setups }
\label{sec:dataset_evaluation}

\subsection{Datasets}
\label{sec:dataset}
In our evaluation, we consider a comprehensive set of four open-sources MI datasets from MOABB~\cite{jayaram2018moabb}. This selection includes two small-scale datasets, namely BNCI~\cite{tangermann2012review} (the dataset IIa from BCI Competition 4) and Zhou~\cite{zhou2016fully}, as well as two large datasets, Cho\cite{cho2017eeg} and Physionet\cite{goldberger2000physiobank}, respectively named BNCI2014001, Zhou2016, Cho2017 and PhysionetMI on MOABB. These datasets were chosen due to their frequent utilization in the field of MI classification.

We include both small and large datasets to ensure the robustness of our experiments and ablations. This inclusion provides strong evidence for the reliability and effectiveness of the proposed architecture across varying dataset sizes.

Each dataset has a unique recording setup with different devices and configurations, detailed in table~\ref{tab:datasets}. We exclude specific subjects from Cho (S31, S45, S48) and Physionet (S37, S87, S88, S91, S99, S103) datasets, following precedents like~\cite{kostas2020thinker,sleight2009classification,roots2020fusion}, due to unusable data or missing trials. This exclusion doesn't impact our comparison to state-of-the-art methods, as it is done using the BNCI dataset.

All four datasets used are offline datasets with similar cue-based protocols, maintaining consistent fundamental characteristics despite slight variations in timing protocols. In our experiments, we focus on signals starting from the cue indicating the task to imagine. By incorporating diverse MI datasets varying in size and recording setup, our evaluation ensures a comprehensive assessment of the proposed architecture's performance and generalizability. The use of well-established, widely-used, and open-sourced datasets strengthens the validity and reliability of our experimental findings.


\begin{table}[H]
\footnotesize{
\begin{tabular}{l|llll}
\hline
Dataset                      & BNCI & Zhou & Physionet & Cho \\ \hline
Considered Subjects              &   9     &   4   & 103       &   49  \\
Sessions per subject  &  2      &  3    &   1        &  1    \\ 
Trials per session &   288     & ~150     &   90        &  200   \\ 
Total Trials       &   5184     & 1800     &   9270        &  9880   \\ 
Classes          &   L/R/F/T     &  L/R/F   &    L/R/F/BH/Re   &  L/R   \\ 
EEG electrodes       &   22     & 14    &   64      & 64    \\ 
EOG electrodes              &   3     &  2    &     0     & 0    \\ 
Sampling frequency (Hz)    &  250 & 250 & 160  &  512   \\ 
Trial duration (s)   &   4     &   5   &   3.5        &  3  \\ \hline
\end{tabular}}
 \caption{MI datasets considered. L = Left hand, R = Right hand, F = Feet, Re = Rest, BH = Both hands, T = Tongue.}
 \label{tab:datasets}
\end{table}

To facilitate comparisons with previous work, we perform a deeper evaluation of our model on the BNCI dataset, the BNCI dataset being commonly used as a reference benchmark in the MI community. We perform evaluation in three different setups, elaborated in the following section.

\subsection{Evaluation setups}
\label{sec:evaluation}

\begin{figure}[t]
\centering
\includegraphics[ width=0.9\linewidth]{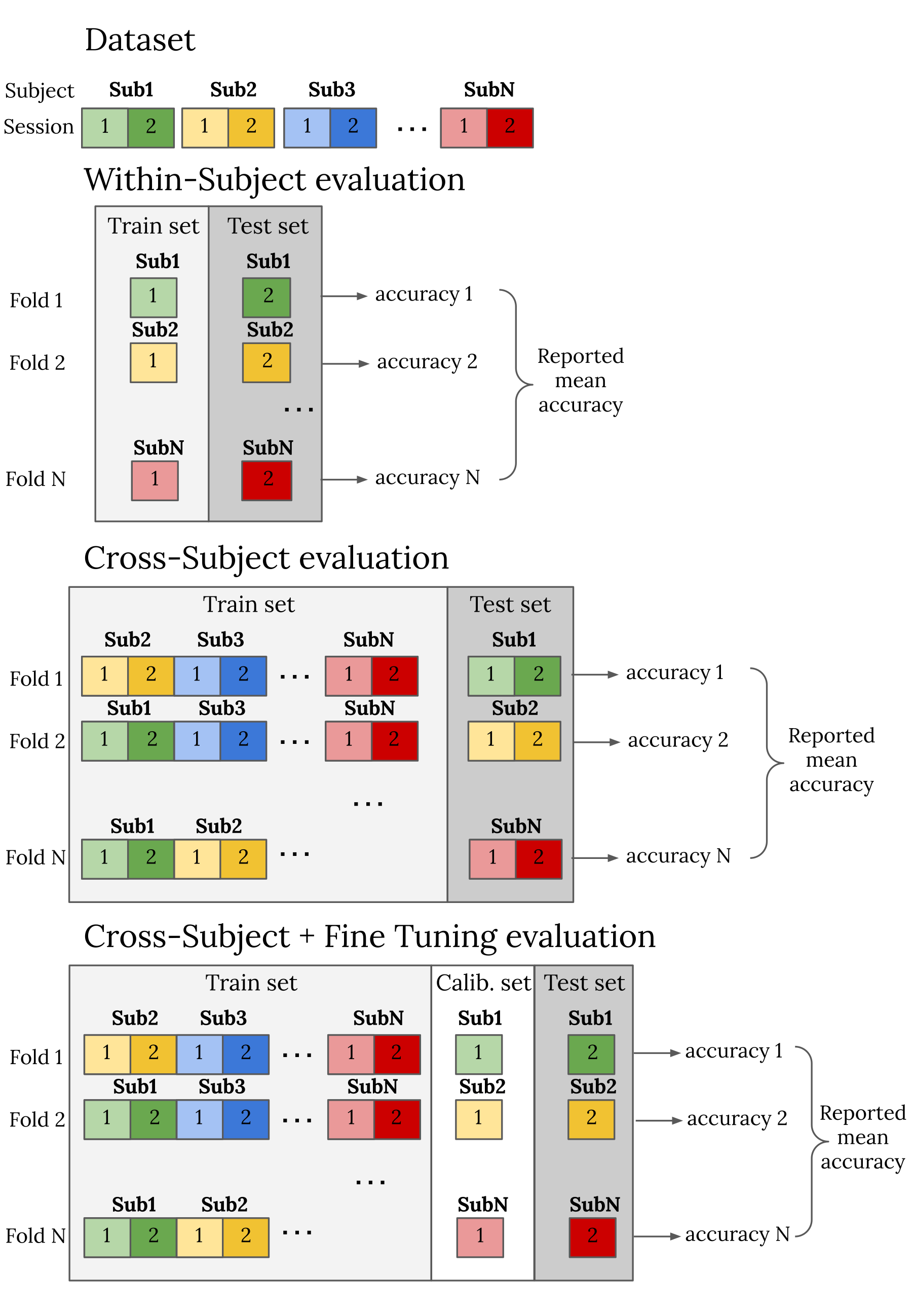}
\caption{Evaluation paradigms under a LOSO approach}
\label{fig:evaluation}
\end{figure}

In our evaluation, we consider three scenarios commonly employed in the MI decoding literature, illustrated in Fig.~\ref{fig:evaluation}.

\subsubsection{Within-Subject (W-S) evaluation}

In MI research, models are mainly assessed within subjects, by training on one part of a subject data and testing on another. Most of Deep Learning models only include this evaluation. In the reported results of the BNCI dataset, we train on the first session and test on the second, producing comparable outcomes to those of the original competition and to the rest of the literature.

\subsubsection{Cross-Subject (C-S) evaluation}

The main advantage of Deep Learning over Machine Learning methods in BCI decoding lies in their ability to leverage information across multiple subjects, showcasing transfer and generalization capabilities. Hence, our primary evaluation focuses on cross-subject assessment. In this paradigm, the model is trained on data from all subjects except one, and its performance is evaluated on the left-out subject's data. This approach provides a robust measure of the model's generalization ability across subjects.

\subsubsection{Cross-Subject with Fine-Tuning (C-S F-T) evaluation}

We also introduce a Cross-subject with Fine-Tuning paradigm, which reflects a realistic usage scenario where a neural network trained on a large dataset is adapted to a specific user. Here, we train the model on data from all subjects except one plus one part of the data of the left-out subject (``Session 1'' in the Fig.~\ref{fig:evaluation}). Then, we evaluate the model's performance on the rest of the data of the left-out subject (``Session 2'' in Fig.~\ref{fig:evaluation}). In the reported results for BNCI, the first session in the left-out subject is used to Fine-Tune, and the second to evaluate. Our Fine-Tuning method consists in two steps. First we train our model on the data from all subjects except one, (like in the Cross-Subject paradigm). In a second step, using the same scheduler and optimizer at the same learning level from the end of the first phase, we Fine-Tune the model on the calibration data for at least 60 additional epochs. The impact of the number of Fine-Tuning epochs on the classification performance is evaluated in Appendix~\ref{app:finetuningepochs}.
In Appendix~\ref{app:FT_vs_MDL}, we compare our fine-tuning method to the Multi-Domain Learning (MDL) method proposed in~\cite{kostas2020thinker}, which consists of a one-step training where the data of the first session of the left-out subject is incorporated with the rest of the training set (data of all other subjects).

In our experiments, we assess the impact of the evaluation paradigm on classification accuracy for the BNCI dataset. Models accuracy in the Within-Subject and in the Cross-Subject with Fine-Tuning paradigms are directly comparable, since both are tested on the second session. However, for the Cross-Subject paradigm, it is not directly comparable with the other two paradigms, since we test on the two sessions available for the left-out subject. To ensure a fair comparison between the three paradigms, we include the result of testing only on the second session for the Cross-Subject paradigm.

\subsubsection{Cross-validation setups}
\label{sec:cvsetups}
To ensure a comprehensive evaluation of our model across all scenarios, we employ cross-validation techniques. For the two smaller datasets, BNCI and Zhou, we utilize the Leave-One-Subject-Out (LOSO) approach. This technique involves leaving out one subject at a time from the dataset for testing/validation, while training the model on the remaining subjects. LOSO enables us to assess the model's generalization performance by evaluating its capability to handle unseen subjects.

For larger datasets, we employ the Leave-Multiple-Subjects-Out (LMSO) technique, with 10 folds. LMSO is an extension of LOSO where multiple subjects are excluded from the training set, allowing us to evaluate the model's performance when faced with scenarios involving multiple missing or excluded subjects. Both LOSO and LMSO serve as effective methodologies for estimating the performance and generalization of Deep Learning models, particularly in our case, where subject-specific variability or subject-wise evaluation is crucial.

By utilizing these cross-validation techniques, we ensure comprehensive evaluation of our model's performance across different evaluation scenarios and dataset sizes. These techniques aid in assessing the model's generalization capabilities, identifying potential issues such as overfitting or subject bias.

\subsubsection{Online evaluation}

To simulate an online use of our model, we add another evaluation, where the model makes predictions on individual trials, one at a time. Consequently, we can't leverage test set data for normalization or alignment. Instead, we compute normalization statistics using training set data. Therefore, for the Within-Subject and Cross-Subject with Fine-Tuning approaches, we compute the statistics of the BNs and EA only on the calibration part of each subject. For the Cross-Subject approach, we exclude EA and  calculate BNs statistics using only data from other subjects.

\section{Results and discussion}
\label{sec:results}

 \subsection{Comparison with the State of the art}
 
\begin{table*}[t]
\caption{Performance comparison to literature methods on BNCI using \textbf{offline} evaluation setups. Standard deviations (stds or $\pm$) represent the variation across subjects.}
\label{tab:sota}
\centering
\begin{threeparttable}
\begin{tabular}{lllll}
\cline{2-5}
&                                             & Within-Subject (W-S)         & Cross-Subject (C-S)               & C-S F-T           \\ \cline{2-5}

\ldelim\{{18.8}{20mm}[\parbox{18.5mm}{\centering Deep\\Learning\\Methods}]
&EEG-SimpleConv                                  & $78.4 \pm 10.6$   & $\mathbf{72.1 \pm 7.3}$                & $\mathbf{86.2\pm6.3}$        \\ 
&EEG Conformer~\cite{song2022eeg}                                  & $\mathbf{78.7 \pm 14.4}$   & ---                & ---        \\ 
&EEG-ITNet~\cite{salami2022eeg}               & $76.7 \pm 11.1$   &  $69.4\pm 8.9$          & $78.7\pm9.4$        \\ 
&EEG-TCNet~\cite{ingolfsson2020eeg} \tnote{1}  & $74.5 \pm 10.1$   &  $65.1\pm10.9$       & $75.8\pm10.2$       \\ 
&CNN-SPDNet~\cite{wilson2022deep}             & $74.2$        &     ---                  &  ---                   \\ 
&Shallow ConvNet~\cite{schirrmeister2017deep} & $73.7$        &    ---                   &  ---                  \\ 
&EEGNet~\cite{lawhern2018eegnet} \tnote{1}    & $73.7 \pm 11.1$   &  $64.0\pm 11.6$       & $73.9\pm 12.2$      \\ 
&EEGNet~\cite{lawhern2018eegnet} \tnote{2}    & ---            &  $66.2\pm 9.9$         & ---      \\ 
&EEG-Inception~\cite{zhang2021eeg} \tnote{1}  & $73.5 \pm 9.1$    &  $66.3\pm 8.7$        & $75.0\pm9.8$        \\ 
&Tensor-SPDNet~\cite{ju2022tensor}            & $73.0$        &   ---                    &  ---                   \\
&Multi-view CNN~\cite{zhang2023multi}         & $72.5 \pm 14.1$   &  ---                     &    ---                 \\ 
&GNN-SPDNet~\cite{ju2022graph}                & $72.0$        &   ---                    &    ---                 \\ 
&Hybrid ConvNet~\cite{schirrmeister2017deep}  & $71.6$        & ---                      &  ---                   \\ 
&Deep ConvNet~\cite{schirrmeister2017deep}    & $70.9$        &  ---                     &  ---                   \\ 
&Residual ConvNet~\cite{schirrmeister2017deep}& $67.7$        &   ---                    &  ---                   \\ 
&DFNN~\cite{li2019densely}                    &   ---                & $64.4$            &   ---                  \\ 
&CCNN~\cite{amin2019deep}                     &   ---                & $55.3$            &   ---                  \\ 
&CMO-CNN~\cite{liu2023compact}                &    ---               & $63.3$            &   ---                  \\ 
&Multi-branch 3D~\cite{zhao2019multi}         &   ---                & $52.2$            &   ---                  \\ 
&TIDNet~\cite{kostas2020thinker}              &    ---               & $65.4$ \tnote{3}  & 77.4 \tnote{3} \\ 
\cline{2-5}
\ldelim\{{2.8}{20mm}[\parbox{18.5mm}{\centering Machine\\Learning\\Methods}]
&CSP+LDA \tnote{2}                                     & $57.7\pm14.9$    &       ---                &   ---                  \\ 
&FBCSP+LDA \tnote{2}                                   & $63.7\pm10.4$    &       ---                &   ---                  \\ 
&TS+LDA \tnote{2}                                     & $65.4\pm12.9$    &       ---                &   ---                  \\ \cline{2-5}

\end{tabular}
\begin{tablenotes}
\hspace{2cm} \item[1] Results reproduced by~\cite{salami2022eeg} \item[2] by us \item[3] 1 subject removed 

  \end{tablenotes}
\end{threeparttable}
\end{table*}

EEG-SimpleConv, under our training routine, is very competitive compared to other state-of-the-art models. In term of classification  performance, EEG-SimpleConv is at least as effective or slightly better than the best models of the literature, and far superior in the Cross-Subject Fine-Tuning paradigm. We note here a real progress in the Fine-Tuning setup, with very promising performances.

In table~\ref{tab:sota}, we report the classification performance of the other state-of-the-art Deep-Learning models. We have chosen to be as exhaustive as possible by including all models that report performance on BNCI in at least one of the three evaluation paradigms that we study. We report the performance indicated in their studies rather than reproducing it. Note that, by doing so we do not really compare models, but rather the models along with their training routines. 

We also report performance of state-of-the-art Machine Learning techniques. For this we compare on the Within-Subject paradigm, as this is where Machine Learning methods are effective. This time we have re-evaluated these baselines ourselves and also made them available on the github repository. We have chosen to compare ourselves to Riemannian methods and CSP-based methods. Among Riemannian methods, we first project the covariance matrices onto the Tangent Space (TS) and then classify using LDA (TS+LDA). For CSP-based methods, we consider a simple CSP combined with an LDA (CSP+LDA), as well as a FBCSP combined with an LDA (FBCSP+LDA), with feature selection performed using the MRMR criterion~\cite{ding2005minimum}. Subject-wise performance are provided in Appendix~\ref{app:ml_subject_details}.

\subsection{The benefits of Cross Subject Transfer Learning}

Through Transfer Learning, EEG-SimpleConv brings a noticeable gain and allows to generalize across subjects. Here, we validate the generalization and transfer capabilities of Deep Learning models in the BCI context. In Table~\ref{tab:evaluation}, we see that in the BNCI dataset, every subject benefits from it, with a gain around 8\% in the Cross-Subject with Fine-Tuning paradigm compared to Within-Subject paradigm. The accuracy gain is way more significant for the ``difficult'' (difficult-to-classify) subjects rather than the ``easy'' (easy-to-classify subjects).

EEG-SimpleConv used directly for transferring, without a calibration (Cross-Subject paradigm) step remains less efficient than if it was trained on the calibration data (Within-Subject paradigm). In the left part of Table~\ref{tab:evaluation}, we see that it still allows a decent classification score, without needing any data from the considered subject, with a drop around only 5\% of accuracy. In this comparison, the drop of accuracy is more significant for the ``easy'' subjects.

The main messages from this comparison are:
\setlist{nolistsep}
\begin{enumerate}
    \item Transferring knowledge, using EEG-SimpleConv, across subjects, allows to classify efficiently.
    \item For a difficult subject (with or without Fine-Tuning), transfer is always efficient, while it is not always the case for an easy subject.
    \item Transferring knowledge with Fine-Tuning on calibration data is helpful for every subject.
\end{enumerate}

\begin{table*}
\caption{EEG-SimpleConv performance on BNCI on various evaluation setups. Stds ($\pm$) on each subject lines represent the variation across runs, while stds on the Average line represent the variation across subjects.}
\label{tab:evaluation}
  \centering
\begin{tabular}{lllll|llll}
\hline
            & \multicolumn{4}{c|}{Offline evaluation}                                        & \multicolumn{4}{c}{Online evaluation}   \\ \hline
             & \multicolumn{1}{c}{W-S} & \multicolumn{2}{c}{C-S}          & \multicolumn{1}{c|}{C-S F-T}       & \multicolumn{1}{c}{W-S}           & \multicolumn{2}{c}{C-S}          & \multicolumn{1}{c}{C-S F-T}    \\ \hline
Test         &  Session 2      & Sess. 1\&2 & Sess. 2             & Sess. 2                 &  Sess. 2        & Sess. 1\&2    & Sess. 2         & Sess. 2   \\ \hline
S0           &  $86.3\pm1.3$   & $78.9\pm1.7$ &  $79.7 \pm 1.5$   &   $89.0\pm1.5$          & $81.0\pm2.0$    &$ 62.5\pm 1.5$ & $64.7\pm 2.0$   & $86.0\pm 1.1$         \\ 
S1           &  $59.1\pm1.1$   & $57.7\pm1.4$ &  $57.3 \pm 1.9$   &   $72.8\pm2.3$          & $54.7\pm2.4$    &$ 50.2\pm 1.0$ & $49.0\pm 2.0$   & $65.8\pm 4.0$       \\ 
S2           &  $91.3\pm0.9$   & $83.0\pm1.1$ &  $85.4 \pm 2.4$   &   $94.1\pm1.3$          & $87.4\pm1.3$    &$ 67.4\pm 2.3$ & $66.8\pm 3.2$   & $87.6\pm 1.6$        \\ 
S3           &  $77.3\pm1.7$   & $66.7\pm1.6$ &  $71.5 \pm 2.2$   &   $89.3\pm0.8$          & $71.9\pm1.9$    &$ 54.5\pm 3.0$ & $57.6\pm 3.2$   & $85.7\pm 1.8$        \\ 
S4           &  $68.3\pm2.3$   & $70.9\pm1.8$ &  $71.0 \pm 2.0$   &   $82.2\pm2.0$          & $44.3\pm5.0$    &$ 56.0\pm 2.7$ & $54.9\pm 4.1$   & $71.5\pm 3.0$        \\ 
S5           &  $68.3\pm1.3$   & $66.6\pm0.7$ &  $65.6 \pm 1.5$   &   $80.3\pm2.3$          & $54.5\pm4.0$    &$ 47.2\pm 2.7$ & $44.8\pm 2.4$   & $68.8\pm 4.5$        \\ 
S6           &  $89.9\pm1.0$   & $75.7\pm1.3$ &  $74.0 \pm 1.9$   &   $92.9\pm0.6$          & $86.5\pm2.9$    &$ 69.1\pm 1.5$ & $70.2\pm 2.0$   & $74.8\pm 1.7$        \\ 
S7           &  $87.2\pm1.0$   & $77.5\pm1.5$ &  $78.1 \pm 1.0$   &   $88.7\pm1.4$          & $84.0\pm1.7$    &$ 60.9\pm 2.6$ & $62.2\pm 2.2$   & $86.7\pm 1.5$        \\ 
S8           &  $77.8\pm0.7$   & $71.6\pm1.6$ &  $71.7 \pm 1.9$   &   $86.1\pm1.2$          & $64.4\pm1.2$    &$ 58.0\pm 3.8$ & $54.3\pm 4.1$   & $81.9\pm 2.8$        \\ \hline 
Average      &  $78.4\pm10.6$  & $72.1\pm7.3$ &  $72.7 \pm 7.7$   &   $86.2\pm6.3$          & $70.0\pm15.1$   &$ 58.4\pm 6.9$& $58.3\pm 7.9$   & $78.8\pm 8.1$        \\ 
+EOG         &  $82.2\pm9.2$   & $79.6\pm6.1$ &  $80.1\pm7.2$      &  $90.2\pm5.5$     \\ \hline
\end{tabular}
\end{table*}

Exploiting EOGs on BNCI, allows a significant gain of between 4 and 8\% depending on the paradigm.

Moving on to an online-like evaluation, as expected we observe in the right part of Table~\ref{tab:evaluation} a loss of classification performance on all paradigms (between 8 to 14\%). Also, the gain made between Within-Subject and Cross-Subject with Fine-Tuning is similar to the same gain in the offline setup (8\% average gain).
Note that, in an online context transferring with calibration phase (C-S F-T) clearly outperforms Machine Learning models (which are trained in a Within-Subject setup, Table~\ref{tab:sota}), while transferring without calibration (C-S) is slightly less efficient than Machine Learning models. Lastly Within-Subject EEG-SimpleConv is slightly more efficient than Machine Learning models.

 
\subsection{Ablation study}
\label{sec:ablation}

From our experiments, we clearly see that Mixup and the Batch Normalisation trick are the most impactful part of our training pipeline.
In this section, we present a comprehensive evaluation of the impact of the various elements of our training pipeline. To gain deeper insights, this study is conducted on four datasets as described in Section~\ref{sec:dataset}. Our assessment focuses on quantifying the influence of each element by systematically removing them one at a time from the pipeline and analyzing their effects on model performance. This ablation is conducted on the cross-subject evaluation paradigm.

\begin{table*}[h]
\caption{Ablation study of our training pipeline ingredients. Stds ($\pm$) represent the variation across subjects. The ``Average Gain'' column is computed by substracting the average accuracy of the full pipeline over each dataset (the first line) to the average accuracy of a given ablation line over each dataset.
}
\label{tab:ablation}
  \centering
\begin{tabular}{llllll}
\hline

                           & \multicolumn{1}{c}{BNCI}                 & \multicolumn{1}{c}{Zhou}                & \multicolumn{1}{c}{Physionet}          & \multicolumn{1}{c}{Cho}                & \multicolumn{1}{c}{Average Gain}      \\ \hline
Full pipeline              & $ 72.1   \pm7.3$  & $81.8 \pm 1.2$    &$ 64.6  \pm 4.4 $ &$ 75.4  \pm 4.6 $ &   \multicolumn{1}{c}{---}                       \\ 
- BN                       & $ 67.1 \pm 8.4 $  & $77.3 \pm 2.9$    &$ 62.8  \pm 4.2 $ &$ 74.5  \pm 5.5 $   & \multicolumn{1}{c}{\textcolor{sr}{$-3.0$}}     \\ 
- EA                       & $ 66.7 \pm 6.5 $  & $80.5 \pm 1.3$    &$ 63.0  \pm 4.6 $ &$ 74.6  \pm 4.6 $   & \multicolumn{1}{c}{\textcolor{sr}{$-2.3$}}     \\ 
- Session                  & $ 70.4 \pm 7.2 $  & $79.9 \pm 1.4$    &     \multicolumn{1}{c}{---}           &  \multicolumn{1}{c}{---}    & \multicolumn{1}{c}{\textcolor{sr}{$-1.8$}}     \\ 
Online                     & $ 58.4 \pm 6.9 $  & $73.1 \pm 6.6$    &$ 61.0  \pm 4.9 $ &$ 73.5  \pm 5.3 $    & \multicolumn{1}{c}{\textcolor{sr}{$-6.8$}}     \\ 
- Mixup                    & $ 69.1 \pm 9.2 $  & $80.1 \pm 0.7$    &$ 60.0  \pm 4.2 $ &$ 73.8  \pm 5.2 $   & \multicolumn{1}{c}{\textcolor{sr}{$-2.7$}}     \\ 
- Reg S                    & $ 71.4 \pm 7.9 $  & $81.8 \pm 1.0$    &$ 63.6  \pm 4.1 $ &$ 74.2  \pm 4.4 $   & \multicolumn{1}{c}{\textcolor{sr}{$-0.72$}}     \\ 
- Everything               & $ 56.4 \pm 9.0 $  & $74.5 \pm 3.6$    &$ 57.4  \pm 4.6 $ &$ 72.6  \pm 4.9 $& \multicolumn{1}{c}{\textcolor{sr}{$-8.3$}}     \\ 
+ EOG                      & $ 79.8 \pm 5.8 $  & $81.9 \pm 1.6$    &     \multicolumn{1}{c}{---}           &    \multicolumn{1}{c}{---}             & \multicolumn{1}{c}{\textcolor{sg}{$+3.9$}}     \\ 
- LMSO                     &     \multicolumn{1}{c}{---}             &     \multicolumn{1}{c}{---}    &  $65.0  \pm 14.3$ & $76.2 \pm 10.3$  & \multicolumn{1}{c}{\textcolor{sg}{$+0.6$}}      \\ \hline
\end{tabular}
\end{table*}

We summarise the results of our evaluation in the Table~\ref{tab:ablation}. To measure performance, we consider two aspects: 1) the classification performance, determined by the average accuracy across all subjects, and 2) the variability of classification performance across subjects, represented by the standard deviation of the model. For the two larger datasets, Cho and Physionet, which contain a significant number of subjects, we utilize the LMSO protocol, described in Table~\ref{sec:cvsetups} to evaluate model performance. Since this protocol provides one standard deviation per fold rather than per subject, we also evaluate the big datasets in LOSO setup (shown at line ``-LMSO''), providing standard deviation per subject and enabling a fair comparison of standard deviations across all datasets. 
In addition to ablating each individual element of the pipeline, we introduce three additional ablation lines. These are: 1) The Online line, where we simultaneously remove at the same time the pipeline elements that requires offline knowledge to simulate an Online evaluation scenario. The removed offline elements are BN, EA and session normalisation. 2) The ``- Everything'' line, serving as a lower limit to measure performance when we remove every additional element of the training pipeline. 3) The EOG line, which investigates the impact of incorporating EOG electrodes for datasets that include them. These ablation lines provide insights into the role and significance of each element in the preprocessing pipeline.

In the last column of the Table~\ref{tab:ablation}, we estimated the average gain provided by each element of the pipeline  by subtracting the average accuracy of the full pipeline over each dataset (the first line) to the average accuracy of the ablation line over each dataset.

\textbf{Impact of Normalisation}: Among the Normalisation elements, the BN trick has the largest impact, but EA and Session also have a significant impact on all datasets. Removing all normalisation elements (the Online line) results in a significant loss of accuracy, of 7\% on average. However, the loss associated with the removal of normalisation elements is more costly for small datasets (containing few subjects) than for large datasets (containing many subjects), with a drop of respectively 12\% versus 2.5\%. We can assume that increasing the number of subjects in a pre-training phase for EEG-SimpleConv allows better generalization in online setups.

We conduct a second ablation only on normalisation, in Appendix~\ref{app:ablation_normalisation}, for that we ablate the normalisation elements of our pipeline, by this time computing all the possible combinations, instead of ablating them one at a time.

\textbf{Impact of Regularisation}: Mixup is the second most important element (after the BN trick). This is justified because EEG-SimpleConv contains a relatively high number of parameters. It would be probably be less impacting, or even damaging to use Mixup for training small networks such as EEGNet. In our case, it is necessary to regularise strongly to prevent overfitting. Regarding the second regularisation, the subject-wise regularisation, it has little or no effect, with a slight gain on large datasets but no gain on small datasets.

\textbf{EOGs}: Adding the EOGs to Zhou doesn't result in a gain like on the BNCI dataset where the gain is important. Understanding exactly why is out of scope of this paper, but we conjecture that it might be due to a difference in their placement.

\textbf{Standard deviations}: We notice very little variability on Zhou's 4 subjects. This may be due to the fact that the subjects had a similar level of expertise in BCI control. Physionet showed very large differences in performance between subjects, with a standard deviation of 14.3\% over around 100 subjects. Lastely, Cho and BNCI show similar deviations between 7\% and 10\%.

Regarding variabilty across runs (instead of across subject), we did not noticed any interesting consistent variation regarding the used pipeline, ranging from 0.7 to 1.7\% for BNCI and Zhou, 0.2 to 0.7\% for Physionet and Cho.

\begin{figure}
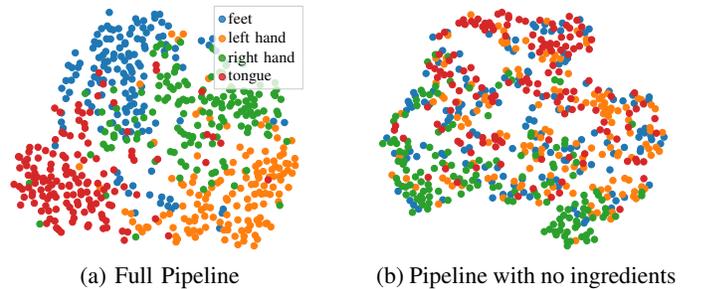

    \centering
    \begin{subfigure}{0.45\linewidth}
        \centering
        \scalebox{0.6}{\input{images/tsne/2True_True_True_True_True_False.tex}}
        \caption{Full Pipeline}
    \end{subfigure}
    \hfill
    \begin{subfigure}{0.45\linewidth}
        \centering
        \scalebox{0.6}{\input{images/tsne/2False_False_False_False_False_False.tex}}
        \caption{Pipeline with no ingredients}
    \end{subfigure}
    \caption{TSNE Embeddings of the Sujbect 2 of BNCI}
    \label{fig:embeddings}
\end{figure}

\textbf{Embeddings}: In addition to the Table~\ref{tab:ablation}, we also include in Fig.~\ref{fig:embeddings} a visualization of the embeddings provided by our model with the full pipeline and the "- Everything" pipeline for Subject 2 of BNCI (the easiest to classify).
Since we are in the cross-subject evaluation paradigm, those embeddings are obtained by passing the whole data of the excluded subject into EEG-SimpleConv. We then extract the representation of this data at the end of the neural network, just before the final classification layer. We then project this representation into a 2D space using TSNE~\cite{ljpvd2008visualizing}. We can clearly see the positive impact of the training pipeline to help EEG-Simpleconv to discriminate adequately the different MI classes. We also provide additional embeddings visualisation  for the \textit{hardest} and the \textit{average} to classify BNCI subject in Appendix~\ref{app:ablation_pipeline}.

\subsection{Inference time evaluation}
\label{sec:time_inference}
EEG-SimpleConv is also very competitive in term of inference time. It has a low latency, comparable to the rest of the models in the literature. This shows that a straightforward architecture, despite having a high number of parameters, can still exhibit remarkably low latency and remain highly competitive and practical for real-world applications.

In Table~\ref{tab:time}, we measure the inference time and number of parameters and compare EEG-SimpleConv with state-of-the-art networks. The reference models were chosen because they are available on Braindecode~\cite{schirrmeister2017deep}. To measure the inference time, we put the network in inference mode and run it 5760 times (\emph{i.e.}, 10 times all the 576 trials available for the first subject of BNCI). We pass the data one by one (not in batch) to simulate its use in an online practical case. We then report the average inference time for processing a single element.

We can see that the Within-Subject configuration of EEG-SimpleConv (where a model is trained only on a subject's data, named EEG-SimpleConv$_w$ in the table) has a lower latency compared to EEGNet, despite having 60 times more parameters. Our findings challenge the conventional assumption that a decrease in the number of parameters will necessarily lead to improved latency and efficiency. Additionally, we validate the fact that achieving a low latency can also be done by simplifying the architecture. The experiments were done on GPU (NVIDIA GeForce RTX 3080 10 Go), as well as on CPU (Intel(R) Xeon(R) W-2223 @ 3.60GHz).

In Table~\ref{tab:time}, we also see that the Cross-Subject version of EEG-SimpleConv (EEG-SimpleConv$_c$) is slower and larger than within-EEG-SimpleConv, because it is way deeper. 

However, the Cross-Subject model still has an inference time comparable to the rest of the reference models. It is worth noticing that in general models are designed to be efficient in the within-subject paradigm, which implies that it requires less parameters to be efficient.

In Appendix~\ref{app:time_size}, we also investigate the effect of varying the size and the inference time of EEG-SimpleConv over its classification performance.

\begin{table}[H]
\centering
\caption{Latency and size comparison to literature methods}
\label{tab:time}
\begin{tabular}{llll}
\hline
                & \multicolumn{2}{c}{Inference time (s)} & Size (p)      \\\hline
                & GPU           & CPU                     &        \\\hline
Shallow ConvNet &  $\mathbf{2.65e^{-04}}$ & $7.68e^{-04}$                    & \multicolumn{1}{l}{$38404$}    \\
EEG-SimpleConv$_w$&   $3.60e^{-04}$ & $\mathbf{5.38e^{-04}}$                    & \multicolumn{1}{l}{$245654$}    \\
EEGNet       &  $4.82e^{-04}$ & $1.17e^{-03}$                    & \multicolumn{1}{l}{$\mathbf{1972}$}    \\
EEG-SimpleConv$_c$ &  $1.05e^{-03}$ & $2.56e^{-03}$                    & \multicolumn{1}{l}{$2462246$}     \\
TIDNet          &  $1.06e^{-03}$ & $2.38e^{-03}$                    & \multicolumn{1}{l}{$187028$}     \\
EEG-Inception   &  $1.67e^{-03}$ & $2.54e^{-03}$                    & \multicolumn{1}{l}{$9952$}     \\ \hline
\end{tabular}
\end{table}

\subsection{Perspectives}

Given that the proposed model only uses simple convolutions and a straightforward architecture, we advocate that it can benefit from existing literature in the following domains:
\begin{enumerate}
    \item \textbf{CNN compression}: In CNN compression, authors show the ability to greatly reduce inference time and size of models while maintaining a steady accuracy. Popular methods have been developped such as pruning~\cite{tessier2022rethinking}, quantizing~\cite{courbariaux2015binaryconnect} or distilling~\cite{hinton2015distilling}. This could be employed to embed such a BCI MI decoding model on edge.
    
    \item \textbf{Interpretability}: Extensive methods have been explored to understand and explain the decision-making processes of such models. These include Feature Visualization~\cite{zhang2018interpretable}, Saliency Maps, Grad-CAM~\cite{selvaraju2016grad} or SHAP~\cite{lundberg2017unified}. Such methods could be adapted to interpret features of EEG-SimpleConv.

\end{enumerate}

Future work could include applying our model to other paradigms, where Deep Learning has also started to be explored, such as Steady-State Visual Evoked Potentials (SSVEP), P300, or Emotion decoding. It could also be interesting to evaluate its performances in real-online scenarios.

\section{conclusion}

We introduced EEG-SimpleConv, a competitive, yet simple, state-of-the-art deep learning model for MI decoding in BCI. Our training pipeline allows EEG-SimpleConv in offline evaluation setups to be as good (in Within-Subject and Cross-Subject) or far better (in Cross-Subject with Fine-Tuning) than the best models in the literature, while having a competitive latency.
Our results question the need for ad-hoc deep learning architectures for MI decoding. We also evaluate the direct transfer capabilities of EEG-SimpleConv on several MI datasets under offline evaluation. In online evaluation setups, transfer learning with calibration (Cross-Subject with Fine-Tuning) seems very promising and shows improved results compared to standard Machine Learning approaches. The proposed training pipeline is composed of simple ingredients, among which Mixup (due to the size of our network) and a Batch Normalisation trick seem to be the most effective.
We share the architecture implementation as well as the code to reproduce all experiments in this paper, hoping that EEG-SimpleConv can become a new reliable baseline to compete with or build upon.

\bibliographystyle{IEEEbib}
\bibliography{refs}


\pagebreak
\appendices
 
\pagebreak
In this appendices, we provide additional results we obtain during our experiments. We also include, in Appendix~\ref{app:tune_hyperparam}, guidelines on how to choose the right parameters for the use of EEG-SimpleConv regarding your needs.

\section{Additional Details about Machine Learning models subject-wise performance }
\label{app:ml_subject_details}

In Table~\ref{tab:app_ML}, we include each subject performance for the Machine learning baselines we considered. The most efficient approach is the Riemannian approach, closely followed by the FBCSP. We can see that the pattern of good and bad subjects is the same as the one observed when using EEG-SimpleConv. As stated in the Section~\label{sec:results}, EEG-SimpleConv using offline evaluation outperforms Machine learning methods, but it is not the case while using online evluation. In online evaluation only EEG-SimpleConv outperforms Machine learning methods only in the Fine-Tuning scenario.

\begin{table}[H]
\caption{Subjects performances using Machine learning baselines performances on BNCI. Stds (±) represent the variation across subject}
\label{tab:app_ML}
\centering
\begin{tabular}{llll}
\hline
     & \multicolumn{1}{c}{TS + LDA}    & \multicolumn{1}{c}{CSP+LDA}      & \multicolumn{1}{c}{FBCSP+LDA}   \\ \hline
S0   & $77.4$       & $68.1$        & $75.4$       \\  
S1   & $52.8$       & $52.8$        & $56.3$       \\  
S2   & $84.0$       & $73.6$        & $78.8$       \\  
S3   & $60.1$       & $53.1$        & $64.6$       \\ 
S4   & $47.9$       & $27.8$        & $50.7$       \\  
S5   & $49.3$       & $42.4$        & $47.2$       \\  
S6   & $64.2$       & $55.6$        & $74.3$       \\  
S7   & $74.0$       & $71.5$        & $63.2$       \\  
S8   & $78.8$       & $74.3$        & $63.2$       \\  \hline
mean & $65.4\pm12.9$& $57.7\pm14.9$ & $63.7\pm10.4$ \\  \hline
\end{tabular}
\end{table}

\section{Additional Details for the Cross-Subject with Fine-Tuning paradigm}

\subsection{Impact of the number of Fine-Tuning epochs}
\label{app:finetuningepochs}

\begin{figure}[H]
\centering
\scalebox{0.8}{
\begin{tikzpicture}

\definecolor{darkgray176}{RGB}{176,176,176}
\definecolor{steelblue31119180}{RGB}{31,119,180}

\begin{axis}[
tick align=outside,
tick pos=left,
x grid style={darkgray176},
xlabel={Number of epochs},
xmin=-3.45, xmax=94.45,
xtick style={color=black},
y grid style={darkgray176},
ylabel={Accuracy (\%)},
ymin=71.4174382716049, ymax=87.1195987654321,
ytick style={color=black}
]
\addplot [semithick, steelblue31119180, mark=*, mark size=3, mark options={solid}]
table {%
1 72.1311728395062
2 73.8827160493827
3 75.7654320987654
4 76.7608024691358
5 78.195987654321
6 79.4305555555556
7 80.3410493827161
8 81.2438271604938
9 81.7608024691358
10 82.162037037037
15 83.6126543209877
17 83.6358024691358
20 84.0138888888889
25 84.5077160493827
30 85.2253086419753
45 85.8425925925926
50 85.8888888888889
60 86.2438271604938
70 86.2283950617284
80 86.1589506172839
90 86.4058641975308
};
\end{axis}

\end{tikzpicture}
}
\caption{Impact of the number Fine-Tuning epochs on the classification performance on BNCI using the C.S.+F.T. paradigm}
\label{fig:figfinetuneepoch}
\end{figure}
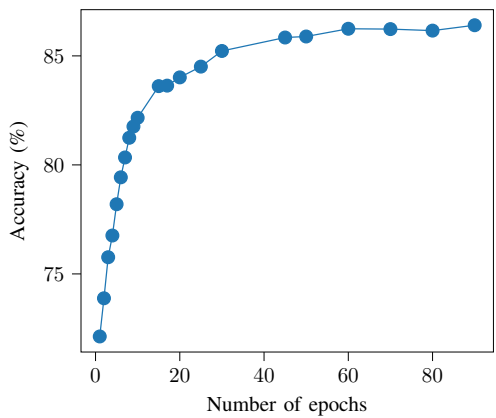

In Fig.~\ref{fig:figfinetuneepoch}, we can see that EEG-SimpleConv seems to require around 60 epochs to be properly finetuned on the BNCI dataset.

\subsection{Fine tuning MDL subjects details}
\label{app:FT_vs_MDL}

In Table~\ref{tab:MDL_details}, we add details regarding each subject performance. The pattern of good and bad subjects remains similar.

\begin{table}[H]
\centering
\caption{Our Fine-Tuning method vs MDL, subject details. Stds (±) on each subject lines represent the variation across runs, while stds on the Average line represent the variation across subject}
\label{tab:MDL_details}
\small{
\begin{tabular}{lll}
\hline
         & \multicolumn{1}{c}{C.S.+F.T.}          & \multicolumn{1}{c}{MDL}            \\ \hline
S0        &   $89.0\pm1.5$   &  $87.4\pm1.0$    \\
S1        &   $72.8\pm2.3$   &  $71.7\pm1.2$   \\ 
S2        &   $94.1\pm1.3$   &  $92.2\pm1.5$   \\ 
S3        &   $89.3\pm0.8$   &  $82.0\pm0.9$   \\ 
S4        &   $82.2\pm2.0$   &  $75.9\pm0.3$   \\ 
S5        &   $80.3\pm2.3$   &  $76.1\pm1.1$   \\ 
S6        &   $92.9\pm0.6$   &  $84.8\pm1.9$   \\ 
S7        &   $88.7\pm1.4$   &  $84.8\pm1.3$   \\ 
S8        &   $86.1\pm1.2$   &  $81.3\pm2.3$   \\ 
mean      &   $86.2\pm6.3$   &  $81.9\pm6.2$   \\ 
+EOG      &   $90.2\pm5.5$   &  $88.4\pm6.4$  \\ \hline

\end{tabular}}
\end{table}

\section{Additional Details about our Ablation study}

\subsection{Normalisation}

In Table~\ref{tab:normalisation}, we report the ablation over all possibilities of the normalisation items.
Overall, as seen in Section~\ref{sec:ablation} normalizing has a big impact on classification, with a boost of 14\%. From the detailed results the observation is similar to what as been previously seen, BN has the bigger impact followed by EA.
Performing a Zscore does not add much, even compared with no normalisation at all. It could make sense because EEG data are from the same dataset and are already with a similar scale. 

\label{app:ablation_normalisation}
\begin{table}[H]
\centering
\caption{Normalisation pipeline full ablation}
\label{tab:normalisation}
\begin{tabular}{ccccc}
\hline
Session & BN    & EA    & Z0      & Accuracy    \\ \hline
\cmark    & \cmark  & \cmark  & \cmark      & $72.1$ \\ 
\cmark    & \cmark  & \cmark  & \xmark      & $71.9$ \\ 
\cmark    & \cmark  & \xmark & \cmark       & $67.7$ \\ 
\cmark    & \cmark  & \xmark & \xmark       & $67.5$ \\ 
\cmark    & \xmark & \cmark  & \cmark       & $67.1$ \\ 
\cmark    & \xmark & \cmark  & \xmark       & $64.0$ \\ 
\cmark    & \xmark & \xmark & \cmark        & $58.4$ \\ 
\cmark    & \xmark & \xmark & \xmark        & $58.3$ \\ 
\xmark   & \cmark  & \cmark  & \cmark       & $71.4$ \\ 
\xmark   & \cmark  & \cmark  & \xmark       & $70.4$ \\ 
\xmark   & \cmark  & \xmark & \cmark        & $67.1$ \\ 
\xmark   & \cmark  & \xmark & \xmark        & $68.1$ \\ 
\xmark   & \xmark & \cmark  & \cmark        & $63.6$ \\ 
\xmark   & \xmark & \cmark  & \xmark        & $63.5$ \\ 
\xmark   & \xmark & \xmark & \cmark         & $58.2$ \\ 
\xmark   & \xmark & \xmark & \xmark         & $58.6$ \\ \hline
\end{tabular}
\end{table}

\subsection{Additionnal embeddings}
\label{app:ablation_pipeline}

\begin{figure}[H]
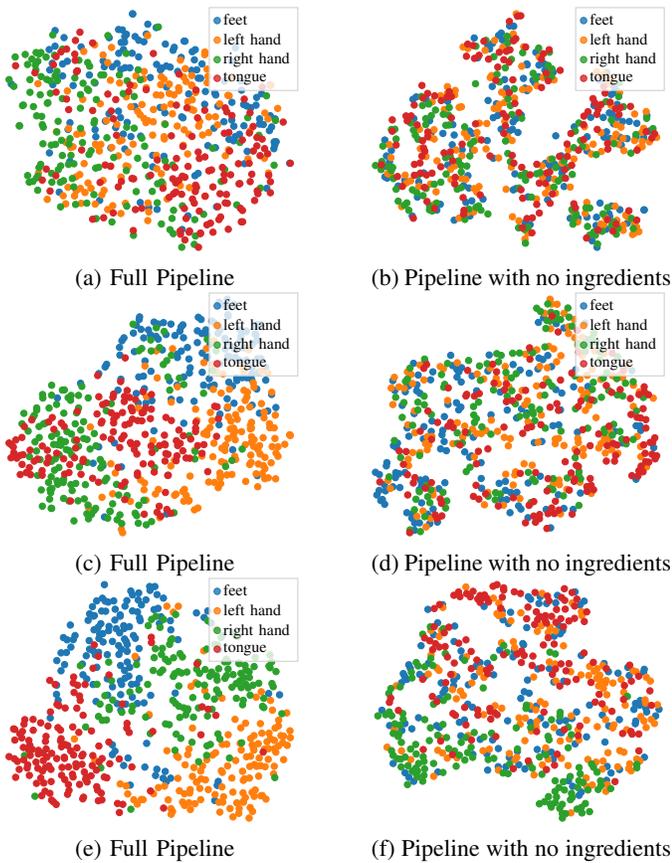

    \centering
    \begin{subfigure}{0.45\linewidth}
        \centering
        \scalebox{0.6}{\input{images/tsne/1True_True_True_True_True_False.tex}}
        \caption{Full Pipeline}
    \end{subfigure}
    \hfill
    \begin{subfigure}{0.45\linewidth}
        \centering
        \scalebox{0.6}{\input{images/tsne/1False_False_False_False_False_False.tex}}
        \caption{Pipeline with no ingredients}
    \end{subfigure}
    
    \begin{subfigure}{0.45\linewidth}
        \centering
        \scalebox{0.6}{\input{images/tsne/6True_True_True_True_True_False.tex}}
        \caption{Full Pipeline}
    \end{subfigure}
    \hfill
    \begin{subfigure}{0.45\linewidth}
        \centering
        \scalebox{0.6}{\input{images/tsne/6False_False_False_False_False_False.tex}}
        \caption{Pipeline with no ingredients}
    \end{subfigure}
    
    \begin{subfigure}{0.45\linewidth}
        \centering
        \scalebox{0.6}{\input{images/tsne/2True_True_True_True_True_False.tex}}
        \caption{Full Pipeline}
    \end{subfigure}
    \hfill
    \begin{subfigure}{0.45\linewidth}
        \centering
        \scalebox{0.6}{\input{images/tsne/2False_False_False_False_False_False.tex}}
        \caption{Pipeline with no ingredients}
    \end{subfigure}
    \caption{TSNE Embeddings. From top to bottom: (a,b) the ``hardest'', (c,d) the ``average''  and (e,f) the ``easiest'' BNCI subject to classify}
    \label{fig:app_tnse}
\end{figure}

In Fig.~\ref{fig:app_tnse}, we provide visualisations of EEG-Simpleconv embeddings right before the classification for three additional subjects.


\section{How to choose the right hyperameter for EEG-SimpleConv}
\label{app:tune_hyperparam}
In this section, we investigate the process of selecting the optimal parameters for our architecture and preprocessing. We hope that this section might facilitate the use of our model. To achieve this, we conducted a series of experiments on the four datasets included in our study, systematically varying different hyperparameters. The results of these experiments are summarized in Table~\ref{tab:hyperparameter}, which provides a concise overview of the recommended parameter ranges for conducting further experiments, related to the specific evaluation paradigm.

\begin{table}[H]
\caption{Hyperparameter Search space }
\label{tab:hyperparameter}
\centering
\begin{tabular}{ccc}
\hline
                      & Within      & Cross        \\\hline
Kernel Size           & ${[}12-17{]}$ & ${[}5-8{]}$    \\
Length ($W$) &   ${[}64-144{]}$          & ${[}64-144{]}$ \\
Depth ($K$)        & $1$           & ${[}2-4{]}$   \\
Sampling Freq. (Hz.)        &    ${[}70-100{]}$          & ${[}50-80{]}$  \\
High-Pass Freq. (Hz.)      & $0.5$         & $0.5$      \\   \hline
\end{tabular}
\end{table}

The table shows the ideal ranges for selecting the following parameters:
\setlist{nolistsep}
\begin{enumerate}[noitemsep]
    \item The size of the convolution kernels employed in the architecture.
    \item $W$, the width of the network, denoted by the number of features maps in the first convolutional layer.
    \item $K$, the depth of the network, related to the number of convolutional layers.
    \item Data pre-processing aspects, including the resampling frequency and the high-pass filter frequency.    
\end{enumerate}

We notice that, if the EEGSimpleconv is too big, whether in term of Depth or Width, it leads to a drop in performance, indicating potential overfitting.

Also, there is an important difference in the dimensions of EEG-Simpleconv with respect to the evaluation paradigm. Under a Within-Subject setup, the optimal model is less deep (with around 3 convolutions instead of 7-11), and requires way bigger kernels (of size 15 instead of 5).

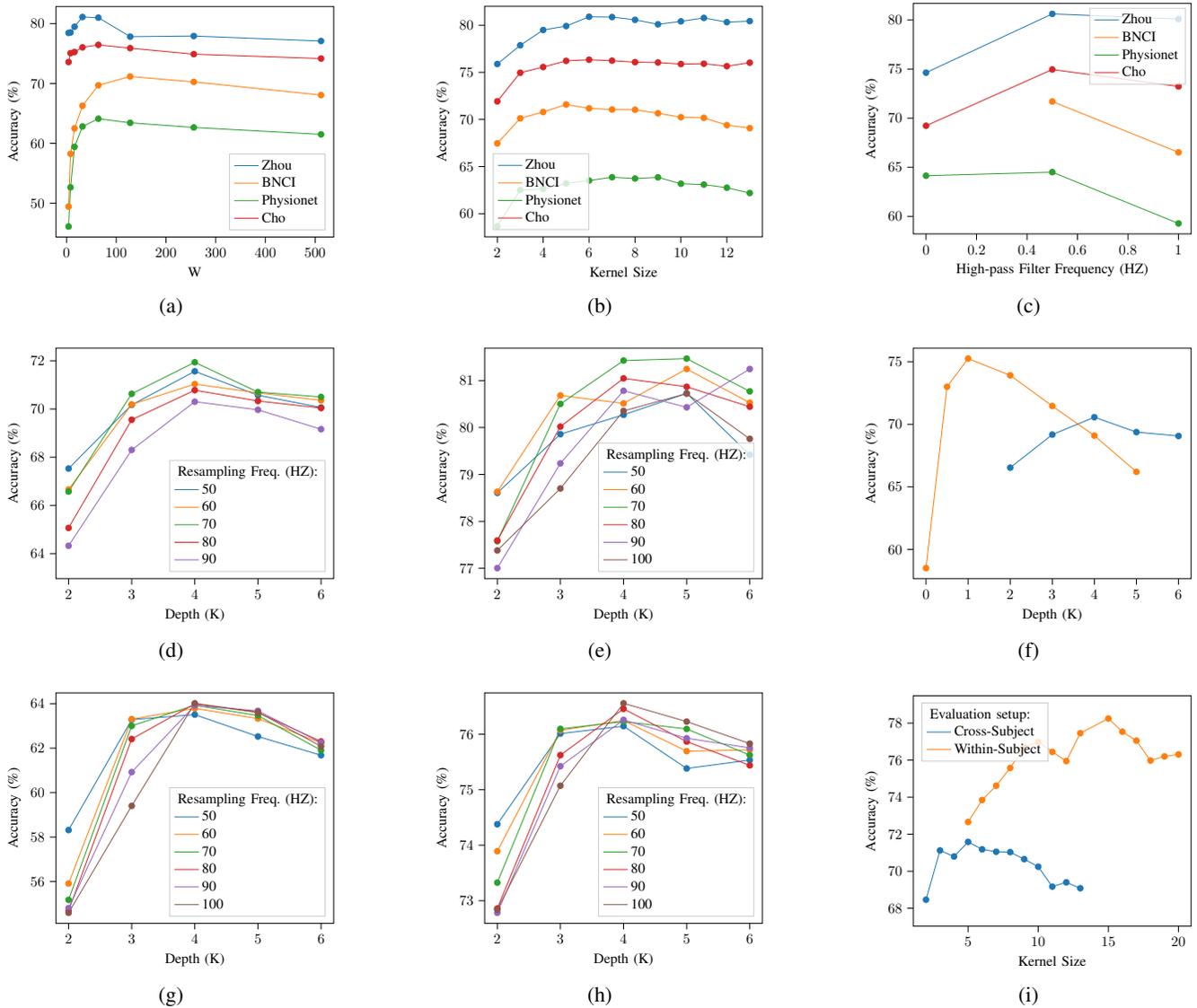
\begin{figure*}
    \begin{subfigure}{0.3\linewidth}
        \centering
        \scalebox{0.6}{
\begin{tikzpicture}

\definecolor{crimson2143940}{RGB}{214,39,40}
\definecolor{darkgray176}{RGB}{176,176,176}
\definecolor{darkorange25512714}{RGB}{255,127,14}
\definecolor{forestgreen4416044}{RGB}{44,160,44}
\definecolor{lightgray204}{RGB}{204,204,204}
\definecolor{steelblue31119180}{RGB}{31,119,180}

\begin{axis}[
legend cell align={left},
legend style={
  fill opacity=0.8,
  draw opacity=1,
  text opacity=1,
  at={(0.97,0.03)},
  anchor=south east,
  draw=lightgray204
},
tick align=outside,
tick pos=left,
x grid style={darkgray176},
xlabel={W},
xmin=-21.4, xmax=537.4,
xtick style={color=black},
y grid style={darkgray176},
ylabel={Accuracy (\%)},
ymin=44.3567047423475, ymax=82.8507641967106,
ytick style={color=black}
]
\addplot [draw=steelblue31119180, fill=steelblue31119180, forget plot, mark=*, only marks]
table{%
x  y
4 78.4445001344275
8 78.5234770789341
16 79.4765937927768
32 81.1010342215123
64 80.9906061994763
128 77.8221449408426
256 77.927486215
512 77.0894582427019
};
\addplot [draw=darkorange25512714, fill=darkorange25512714, forget plot, mark=*, only marks]
table{%
x  y
4 49.4367283950617
8 58.2484567901235
16 62.4891975308642
32 66.2854938271605
64 69.7037037037037
128 71.1658950617284
256 70.2631172839506
512 68.0679012345679
};
\addplot [draw=forestgreen4416044, fill=forestgreen4416044, forget plot, mark=*, only marks]
table{%
x  y
4 46.1064347175458
8 52.6569397680509
16 59.4225589225589
32 62.8047138047138
64 64.1086793864572
128 63.4287317620651
256 62.6505798728021
512 61.4960718294052
};
\addplot [draw=crimson2143940, fill=crimson2143940, forget plot, mark=*, only marks]
table{%
x  y
4 73.5869135802469
8 75.0681893004115
16 75.2621399176955
32 76.0337448559671
64 76.4543621399177
128 75.9011934156378
256 74.8979835390946
512 74.1640740740741
};
\addplot [semithick, steelblue31119180]
table {%
4 78.4445001344275
8 78.5234770789341
16 79.4765937927768
32 81.1010342215123
64 80.9906061994763
128 77.8221449408426
256 77.927486215
512 77.0894582427019
};
\addlegendentry{Zhou}
\addplot [semithick, darkorange25512714]
table {%
4 49.4367283950617
8 58.2484567901235
16 62.4891975308642
32 66.2854938271605
64 69.7037037037037
128 71.1658950617284
256 70.2631172839506
512 68.0679012345679
};
\addlegendentry{BNCI}
\addplot [semithick, forestgreen4416044]
table {%
4 46.1064347175458
8 52.6569397680509
16 59.4225589225589
32 62.8047138047138
64 64.1086793864572
128 63.4287317620651
256 62.6505798728021
512 61.4960718294052
};
\addlegendentry{Physionet}
\addplot [semithick, crimson2143940]
table {%
4 73.5869135802469
8 75.0681893004115
16 75.2621399176955
32 76.0337448559671
64 76.4543621399177
128 75.9011934156378
256 74.8979835390946
512 74.1640740740741
};
\addlegendentry{Cho}
\end{axis}

\end{tikzpicture}}
        \caption{}
    \end{subfigure}
    \hfill
    \begin{subfigure}{0.3\linewidth}
        \centering
        \scalebox{0.6}{
\begin{tikzpicture}

\definecolor{crimson2143940}{RGB}{214,39,40}
\definecolor{darkgray176}{RGB}{176,176,176}
\definecolor{darkorange25512714}{RGB}{255,127,14}
\definecolor{forestgreen4416044}{RGB}{44,160,44}
\definecolor{lightgray204}{RGB}{204,204,204}
\definecolor{steelblue31119180}{RGB}{31,119,180}

\begin{axis}[
legend cell align={left},
legend style={
  fill opacity=0.8,
  draw opacity=1,
  text opacity=1,
  at={(0.03,0.03)},
  anchor=south west,
  draw=lightgray204
},
tick align=outside,
tick pos=left,
x grid style={darkgray176},
xlabel={Kernel Size},
xmin=1.45, xmax=13.55,
xtick style={color=black},
y grid style={darkgray176},
ylabel={Accuracy (\%)},
ymin=57.5511346687228, ymax=82.0105340967396,
ytick style={color=black}
]
\addplot [draw=steelblue31119180, fill=steelblue31119180, forget plot, mark=*, only marks]
table{%
x  y
2 75.8826258586079
3 77.8708487573747
4 79.4968950262797
5 79.9140052131145
6 80.8987432136479
7 80.8691827400993
8 80.5755166545062
9 80.1023516265883
10 80.4115684659834
11 80.7779751579755
12 80.32597810543
13 80.438565001091
};
\addplot [draw=darkorange25512714, fill=darkorange25512714, forget plot, mark=*, only marks]
table{%
x  y
2 67.4614197530864
3 70.1234567901235
4 70.795524691358
5 71.5864197530864
6 71.1813271604938
7 71.054012345679
8 71.0308641975309
9 70.6527777777778
10 70.2438271604938
11 70.170524691358
12 69.3981481481482
13 69.0817901234568
};
\addplot [draw=forestgreen4416044, fill=forestgreen4416044, forget plot, mark=*, only marks]
table{%
x  y
2 58.6629255518144
3 62.5153385708941
4 62.6249532360643
5 63.2171717171717
6 63.5188926300037
7 63.8707444818556
8 63.7328843995511
9 63.8567153011597
10 63.1801346801347
11 63.0909090909091
12 62.7626262626263
13 62.1994014216236
};
\addplot [draw=crimson2143940, fill=crimson2143940, forget plot, mark=*, only marks]
table{%
x  y
2 71.9263374485597
3 74.9618518518519
4 75.5655967078189
5 76.2222222222222
6 76.3407818930041
7 76.2430041152263
8 76.0842386831276
9 76.0458436213992
10 75.8832921810699
11 75.9186008230453
12 75.6579835390947
13 76.0302880658436
};
\addplot [semithick, steelblue31119180]
table {%
2 75.8826258586079
3 77.8708487573747
4 79.4968950262797
5 79.9140052131145
6 80.8987432136479
7 80.8691827400993
8 80.5755166545062
9 80.1023516265883
10 80.4115684659834
11 80.7779751579755
12 80.32597810543
13 80.438565001091
};
\addlegendentry{Zhou}
\addplot [semithick, darkorange25512714]
table {%
2 67.4614197530864
3 70.1234567901235
4 70.795524691358
5 71.5864197530864
6 71.1813271604938
7 71.054012345679
8 71.0308641975309
9 70.6527777777778
10 70.2438271604938
11 70.170524691358
12 69.3981481481482
13 69.0817901234568
};
\addlegendentry{BNCI}
\addplot [semithick, forestgreen4416044]
table {%
2 58.6629255518144
3 62.5153385708941
4 62.6249532360643
5 63.2171717171717
6 63.5188926300037
7 63.8707444818556
8 63.7328843995511
9 63.8567153011597
10 63.1801346801347
11 63.0909090909091
12 62.7626262626263
13 62.1994014216236
};
\addlegendentry{Physionet}
\addplot [semithick, crimson2143940]
table {%
2 71.9263374485597
3 74.9618518518519
4 75.5655967078189
5 76.2222222222222
6 76.3407818930041
7 76.2430041152263
8 76.0842386831276
9 76.0458436213992
10 75.8832921810699
11 75.9186008230453
12 75.6579835390947
13 76.0302880658436
};
\addlegendentry{Cho}
\end{axis}

\end{tikzpicture}}
        \caption{}
    \end{subfigure}
    \hfill
    \begin{subfigure}{0.3\linewidth}
        \centering
        \scalebox{0.6}{
\begin{tikzpicture}

\definecolor{crimson2143940}{RGB}{214,39,40}
\definecolor{darkgray176}{RGB}{176,176,176}
\definecolor{darkorange25512714}{RGB}{255,127,14}
\definecolor{forestgreen4416044}{RGB}{44,160,44}
\definecolor{lightgray204}{RGB}{204,204,204}
\definecolor{steelblue31119180}{RGB}{31,119,180}

\begin{axis}[
legend cell align={left},
legend style={fill opacity=0.8, draw opacity=1, text opacity=1, draw=lightgray204},
tick align=outside,
tick pos=left,
x grid style={darkgray176},
xlabel={High-pass Filter Frequency (HZ)},
xmin=-0.05, xmax=1.05,
xtick style={color=black},
y grid style={darkgray176},
ylabel={Accuracy (\%)},
ymin=58.2117961194479, ymax=81.6792814915933,
ytick style={color=black}
]
\addplot [draw=steelblue31119180, fill=steelblue31119180, forget plot, mark=*, only marks]
table{%
x  y
0 74.6324020567056
0.5 80.6125776110413
1 80.0899148118843
};
\addplot [draw=darkorange25512714, fill=darkorange25512714, forget plot, mark=*, only marks]
table{%
x  y
0.5 71.6944
1 66.51
};
\addplot [draw=forestgreen4416044, fill=forestgreen4416044, forget plot, mark=*, only marks]
table{%
x  y
0 64.1343
0.5 64.5
1 59.2785
};
\addplot [draw=crimson2143940, fill=crimson2143940, forget plot, mark=*, only marks]
table{%
x  y
0 69.2181069958848
0.5 74.952304526749
1 73.2255144032922
};
\addplot [semithick, steelblue31119180]
table {%
0 74.6324020567056
0.5 80.6125776110413
1 80.0899148118843
};
\addlegendentry{Zhou}
\addplot [semithick, darkorange25512714]
table {%
0.5 71.6944
1 66.51
};
\addlegendentry{BNCI}
\addplot [semithick, forestgreen4416044]
table {%
0 64.1343
0.5 64.5
1 59.2785
};
\addlegendentry{Physionet}
\addplot [semithick, crimson2143940]
table {%
0 69.2181069958848
0.5 74.952304526749
1 73.2255144032922
};
\addlegendentry{Cho}
\end{axis}

\end{tikzpicture}}
        \caption{}
    \end{subfigure}
    \\~\\
    \begin{subfigure}{0.3\linewidth}
        \centering
        \scalebox{0.6}{
\begin{tikzpicture}

\definecolor{crimson2143940}{RGB}{214,39,40}
\definecolor{darkgray176}{RGB}{176,176,176}
\definecolor{darkorange25512714}{RGB}{255,127,14}
\definecolor{forestgreen4416044}{RGB}{44,160,44}
\definecolor{lightgray204}{RGB}{204,204,204}
\definecolor{mediumpurple148103189}{RGB}{148,103,189}
\definecolor{sienna1408675}{RGB}{140,86,75}
\definecolor{steelblue31119180}{RGB}{31,119,180}

\begin{axis}[
legend cell align={left},
legend style={
  fill opacity=0.8,
  draw opacity=1,
  text opacity=1,
  at={(0.97,0.03)},
  anchor=south east,
  draw=lightgray204
},
tick align=outside,
tick pos=left,
unbounded coords=jump,
x grid style={darkgray176},
xlabel={Depth (K)},
xmin=1.8, xmax=6.2,
xtick style={color=black},
y grid style={darkgray176},
ylabel={Accuracy (\%)},
ymin=62.9637345679012, ymax=72.5251543209876,
ytick style={color=black}
]
\addlegendimage{empty legend}
\addlegendentry{\hspace{-.6cm}Resampling Freq. (HZ):}
\addplot [draw=steelblue31119180, fill=steelblue31119180, forget plot, mark=*, only marks]
table{%
x  y
2 67.5277777777778
3 70.170524691358
4 71.5632716049383
5 70.5711419753086
6 70.0625
};
\addplot [draw=darkorange25512714, fill=darkorange25512714, forget plot, mark=*, only marks]
table{%
x  y
2 66.6597222222222
3 70.1898148148148
4 71.0347222222222
5 70.6714506172839
6 70.3595679012346
};
\addplot [draw=forestgreen4416044, fill=forestgreen4416044, forget plot, mark=*, only marks]
table{%
x  y
2 66.5717592592593
3 70.6304012345679
4 71.9382716049383
5 70.7010802469136
6 70.4984567901235
};
\addplot [draw=crimson2143940, fill=crimson2143940, forget plot, mark=*, only marks]
table{%
x  y
2 65.0663580246914
3 69.5532407407407
4 70.7800925925926
5 70.3325617283951
6 70.0393518518519
};
\addplot [draw=mediumpurple148103189, fill=mediumpurple148103189, forget plot, mark=*, only marks]
table{%
x  y
2 64.3256172839506
3 68.2993827160494
4 70.3016975308642
5 69.9660493827161
6 69.1597222222222
};
\addplot [draw=none, draw=sienna1408675, fill=sienna1408675, forget plot, mark=*]
table{%
x  y
0 -0.5
0.13260155 -0.5
0.259789935392427 -0.447316845794121
0.353553390593274 -0.353553390593274
0.447316845794121 -0.259789935392427
0.5 -0.13260155
0.5 0
0.5 0.13260155
0.447316845794121 0.259789935392427
0.353553390593274 0.353553390593274
0.259789935392427 0.447316845794121
0.13260155 0.5
0 0.5
-0.13260155 0.5
-0.259789935392427 0.447316845794121
-0.353553390593274 0.353553390593274
-0.447316845794121 0.259789935392427
-0.5 0.13260155
-0.5 0
-0.5 -0.13260155
-0.447316845794121 -0.259789935392427
-0.353553390593274 -0.353553390593274
-0.259789935392427 -0.447316845794121
-0.13260155 -0.5
0 -0.5
0 -0.5
};
\addplot [semithick, steelblue31119180]
table {%
2 67.5277777777778
3 70.170524691358
4 71.5632716049383
5 70.5711419753086
6 70.0625
};
\addlegendentry{50}
\addplot [semithick, darkorange25512714]
table {%
2 66.6597222222222
3 70.1898148148148
4 71.0347222222222
5 70.6714506172839
6 70.3595679012346
};
\addlegendentry{60}
\addplot [semithick, forestgreen4416044]
table {%
2 66.5717592592593
3 70.6304012345679
4 71.9382716049383
5 70.7010802469136
6 70.4984567901235
};
\addlegendentry{70}
\addplot [semithick, crimson2143940]
table {%
2 65.0663580246914
3 69.5532407407407
4 70.7800925925926
5 70.3325617283951
6 70.0393518518519
};
\addlegendentry{80}
\addplot [semithick, mediumpurple148103189]
table {%
2 64.3256172839506
3 68.2993827160494
4 70.3016975308642
5 69.9660493827161
6 69.1597222222222
};
\addlegendentry{90}
\addplot [semithick, sienna1408675]
table {%
2 nan
};
\addlegendentry{100}
\end{axis}

\end{tikzpicture}}
        \caption{}
    \end{subfigure}
    \hfill
    \begin{subfigure}{0.3\linewidth}
        \centering
        \scalebox{0.6}{
\begin{tikzpicture}

\definecolor{crimson2143940}{RGB}{214,39,40}
\definecolor{darkgray176}{RGB}{176,176,176}
\definecolor{darkorange25512714}{RGB}{255,127,14}
\definecolor{forestgreen4416044}{RGB}{44,160,44}
\definecolor{lightgray204}{RGB}{204,204,204}
\definecolor{mediumpurple148103189}{RGB}{148,103,189}
\definecolor{sienna1408675}{RGB}{140,86,75}
\definecolor{steelblue31119180}{RGB}{31,119,180}

\begin{axis}[
legend cell align={left},
legend style={
  fill opacity=0.8,
  draw opacity=1,
  text opacity=1,
  at={(0.97,0.03)},
  anchor=south east,
  draw=lightgray204
},
tick align=outside,
tick pos=left,
x grid style={darkgray176},
xlabel={Depth (K)},
xmin=1.8, xmax=6.2,
xtick style={color=black},
y grid style={darkgray176},
ylabel={Accuracy (\%)},
ymin=76.780787485269, ymax=81.6912836069317,
ytick style={color=black}
]
\addlegendimage{empty legend}
\addlegendentry{\hspace{-.6cm}Resampling Freq. (HZ):}

\addplot [draw=steelblue31119180, fill=steelblue31119180, forget plot, mark=*, only marks]
table{%
x  y
2 78.6067325571391
3 79.8583292058011
4 80.2700255281818
5 80.7269165596807
6 79.4170699091432
};
\addplot [draw=darkorange25512714, fill=darkorange25512714, forget plot, mark=*, only marks]
table{%
x  y
2 78.6311155648909
3 80.6833308299444
4 80.5140516202513
5 81.2476967359055
6 80.5287831460261
};
\addplot [draw=forestgreen4416044, fill=forestgreen4416044, forget plot, mark=*, only marks]
table{%
x  y
2 77.5782498443829
3 80.5024730342983
4 81.426808863412
5 81.4680792377652
6 80.7708111559948
};
\addplot [draw=crimson2143940, fill=crimson2143940, forget plot, mark=*, only marks]
table{%
x  y
2 77.5924340830571
3 80.0187569538211
4 81.0470694574158
5 80.8680474093437
6 80.442348922609
};
\addplot [draw=mediumpurple148103189, fill=mediumpurple148103189, forget plot, mark=*, only marks]
table{%
x  y
2 77.0039918544355
3 79.2360151412196
4 80.7827195720796
5 80.4307633133334
6 81.2472411244488
};
\addplot [draw=sienna1408675, fill=sienna1408675, forget plot, mark=*, only marks]
table{%
x  y
2 77.380371611453
3 78.7014975054169
4 80.353619988121
5 80.722823558757
6 79.758114212316
};
\addplot [semithick, steelblue31119180]
table {%
2 78.6067325571391
3 79.8583292058011
4 80.2700255281818
5 80.7269165596807
6 79.4170699091432
};
\addlegendentry{50}
\addplot [semithick, darkorange25512714]
table {%
2 78.6311155648909
3 80.6833308299444
4 80.5140516202513
5 81.2476967359055
6 80.5287831460261
};
\addlegendentry{60}
\addplot [semithick, forestgreen4416044]
table {%
2 77.5782498443829
3 80.5024730342983
4 81.426808863412
5 81.4680792377652
6 80.7708111559948
};
\addlegendentry{70}
\addplot [semithick, crimson2143940]
table {%
2 77.5924340830571
3 80.0187569538211
4 81.0470694574158
5 80.8680474093437
6 80.442348922609
};
\addlegendentry{80}
\addplot [semithick, mediumpurple148103189]
table {%
2 77.0039918544355
3 79.2360151412196
4 80.7827195720796
5 80.4307633133334
6 81.2472411244488
};
\addlegendentry{90}
\addplot [semithick, sienna1408675]
table {%
2 77.380371611453
3 78.7014975054169
4 80.353619988121
5 80.722823558757
6 79.758114212316
};
\addlegendentry{100}
\end{axis}

\end{tikzpicture}}
        \caption{}
    \end{subfigure}
    \hfill
    \begin{subfigure}{0.3\linewidth}
        \centering
        \scalebox{0.6}{
\begin{tikzpicture}

\definecolor{darkgray176}{RGB}{176,176,176}
\definecolor{darkorange25512714}{RGB}{255,127,14}
\definecolor{lightgray204}{RGB}{204,204,204}
\definecolor{steelblue31119180}{RGB}{31,119,180}

\begin{axis}[
legend cell align={left},
legend style={fill opacity=0.8, draw opacity=1, text opacity=1, draw=lightgray204},
tick align=outside,
tick pos=left,
x grid style={darkgray176},
xlabel={Depth (K)},
xmin=-0.3, xmax=6.3,
xtick style={color=black},
y grid style={darkgray176},
ylabel={Accuracy (\%)},
ymin=57.6626543209877, ymax=76.0842592592593,
ytick style={color=black}
]
\addplot [draw=steelblue31119180, fill=steelblue31119180, forget plot, mark=*, only marks]
table{%
x  y
2 66.5277777777778
3 69.170524691358
4 70.5632716049383
5 69.3711419753086
6 69.0625
};
\addplot [draw=darkorange25512714, fill=darkorange25512714, forget plot, mark=*, only marks]
table{%
x  y
0 58.5
0.5 73
1 75.2469135802469
2 73.912037037037
3 71.4583333333333
4 69.0895061728395
5 66.195987654321
};
\addplot [semithick, steelblue31119180]
table {%
2 66.5277777777778
3 69.170524691358
4 70.5632716049383
5 69.3711419753086
6 69.0625
};
\addplot [semithick, darkorange25512714]
table {%
0 58.5
0.5 73
1 75.2469135802469
2 73.912037037037
3 71.4583333333333
4 69.0895061728395
5 66.195987654321
};
\end{axis}

\end{tikzpicture}}
        \caption{}
    \end{subfigure}
    \\~\\
    \begin{subfigure}{0.3\linewidth}
        \centering
        \scalebox{0.6}{
\begin{tikzpicture}

\definecolor{crimson2143940}{RGB}{214,39,40}
\definecolor{darkgray176}{RGB}{176,176,176}
\definecolor{darkorange25512714}{RGB}{255,127,14}
\definecolor{forestgreen4416044}{RGB}{44,160,44}
\definecolor{lightgray204}{RGB}{204,204,204}
\definecolor{mediumpurple148103189}{RGB}{148,103,189}
\definecolor{sienna1408675}{RGB}{140,86,75}
\definecolor{steelblue31119180}{RGB}{31,119,180}

\begin{axis}[
legend cell align={left},
legend style={
  fill opacity=0.8,
  draw opacity=1,
  text opacity=1,
  at={(0.97,0.03)},
  anchor=south east,
  draw=lightgray204
},
tick align=outside,
tick pos=left,
x grid style={darkgray176},
xlabel={Depth (K)},
xmin=1.8, xmax=6.2,
xtick style={color=black},
y grid style={darkgray176},
ylabel={Accuracy (\%)},
ymin=54.1240460157127, ymax=64.4855686494575,
ytick style={color=black}
]
\addlegendimage{empty legend}
\addlegendentry{\hspace{-.6cm}Resampling Freq. (HZ):}
\addplot [draw=steelblue31119180, fill=steelblue31119180, forget plot, mark=*, only marks]
table{%
x  y
2 58.3172465394688
3 63.2803965581743
4 63.5069210624766
5 62.5220725776281
6 61.6793864571642
};
\addplot [draw=darkorange25512714, fill=darkorange25512714, forget plot, mark=*, only marks]
table{%
x  y
2 55.9120838009727
3 63.3022820800598
4 63.7888140665919
5 63.3299663299663
6 62.2332585110363
};
\addplot [draw=forestgreen4416044, fill=forestgreen4416044, forget plot, mark=*, only marks]
table{%
x  y
2 55.1788252899364
3 63.0014964459409
4 63.9298540965208
5 63.4625888514777
6 61.9036662925552
};
\addplot [draw=crimson2143940, fill=crimson2143940, forget plot, mark=*, only marks]
table{%
x  y
2 54.6952861952862
3 62.4079685746352
4 63.9917695473251
5 63.6002618780397
6 62.2979797979798
};
\addplot [draw=mediumpurple148103189, fill=mediumpurple148103189, forget plot, mark=*, only marks]
table{%
x  y
2 54.807519640853
3 60.921062476618
4 63.9145155256266
5 63.678077066966
6 62.2495323606435
};
\addplot [draw=sienna1408675, fill=sienna1408675, forget plot, mark=*, only marks]
table{%
x  y
2 54.5950243172465
3 59.4027310138421
4 64.0145903479237
5 63.6210250654695
6 62.0800598578376
};
\addplot [semithick, steelblue31119180]
table {%
2 58.3172465394688
3 63.2803965581743
4 63.5069210624766
5 62.5220725776281
6 61.6793864571642
};
\addlegendentry{50}
\addplot [semithick, darkorange25512714]
table {%
2 55.9120838009727
3 63.3022820800598
4 63.7888140665919
5 63.3299663299663
6 62.2332585110363
};
\addlegendentry{60}
\addplot [semithick, forestgreen4416044]
table {%
2 55.1788252899364
3 63.0014964459409
4 63.9298540965208
5 63.4625888514777
6 61.9036662925552
};
\addlegendentry{70}
\addplot [semithick, crimson2143940]
table {%
2 54.6952861952862
3 62.4079685746352
4 63.9917695473251
5 63.6002618780397
6 62.2979797979798
};
\addlegendentry{80}
\addplot [semithick, mediumpurple148103189]
table {%
2 54.807519640853
3 60.921062476618
4 63.9145155256266
5 63.678077066966
6 62.2495323606435
};
\addlegendentry{90}
\addplot [semithick, sienna1408675]
table {%
2 54.5950243172465
3 59.4027310138421
4 64.0145903479237
5 63.6210250654695
6 62.0800598578376
};
\addlegendentry{100}
\end{axis}

\end{tikzpicture}}
        \caption{}
    \end{subfigure}
    \hfill
    \begin{subfigure}{0.3\linewidth}
        \centering
        \scalebox{0.6}{
\begin{tikzpicture}

\definecolor{crimson2143940}{RGB}{214,39,40}
\definecolor{darkgray176}{RGB}{176,176,176}
\definecolor{darkorange25512714}{RGB}{255,127,14}
\definecolor{forestgreen4416044}{RGB}{44,160,44}
\definecolor{lightgray204}{RGB}{204,204,204}
\definecolor{mediumpurple148103189}{RGB}{148,103,189}
\definecolor{sienna1408675}{RGB}{140,86,75}
\definecolor{steelblue31119180}{RGB}{31,119,180}

\begin{axis}[
legend cell align={left},
legend style={
  fill opacity=0.8,
  draw opacity=1,
  text opacity=1,
  at={(0.97,0.03)},
  anchor=south east,
  draw=lightgray204
},
tick align=outside,
tick pos=left,
x grid style={darkgray176},
xlabel={Depth (K)},
xmin=1.8, xmax=6.2,
xtick style={color=black},
y grid style={darkgray176},
ylabel={Accuracy (\%)},
ymin=72.5924074074074, ymax=76.7466049382716,
ytick style={color=black}
]
\addlegendimage{empty legend}
\addlegendentry{\hspace{-.6cm}Resampling Freq. (HZ):}

\addplot [draw=steelblue31119180, fill=steelblue31119180, forget plot, mark=*, only marks]
table{%
x  y
2 74.3787242798354
3 76.0117695473251
4 76.1460905349794
5 75.3830041152263
6 75.5349382716049
};
\addplot [draw=darkorange25512714, fill=darkorange25512714, forget plot, mark=*, only marks]
table{%
x  y
2 73.891316872428
3 76.0707818930041
4 76.2475720164609
5 75.694938271605
6 75.7185596707819
};
\addplot [draw=forestgreen4416044, fill=forestgreen4416044, forget plot, mark=*, only marks]
table{%
x  y
2 73.3250205761317
3 76.0969958847736
4 76.2284362139918
5 76.0956790123457
6 75.623621399177
};
\addplot [draw=crimson2143940, fill=crimson2143940, forget plot, mark=*, only marks]
table{%
x  y
2 72.8584
3 75.6238
4 76.458
5 75.8674
6 75.4394
};
\addplot [draw=mediumpurple148103189, fill=mediumpurple148103189, forget plot, mark=*, only marks]
table{%
x  y
2 72.7812345679012
3 75.4238271604938
4 76.258024691358
5 75.9279012345679
6 75.7541563786008
};
\addplot [draw=sienna1408675, fill=sienna1408675, forget plot, mark=*, only marks]
table{%
x  y
2 72.8383127572016
3 75.0724279835391
4 76.5577777777778
5 76.2285596707819
6 75.830987654321
};
\addplot [semithick, steelblue31119180]
table {%
2 74.3787242798354
3 76.0117695473251
4 76.1460905349794
5 75.3830041152263
6 75.5349382716049
};
\addlegendentry{50}
\addplot [semithick, darkorange25512714]
table {%
2 73.891316872428
3 76.0707818930041
4 76.2475720164609
5 75.694938271605
6 75.7185596707819
};
\addlegendentry{60}
\addplot [semithick, forestgreen4416044]
table {%
2 73.3250205761317
3 76.0969958847736
4 76.2284362139918
5 76.0956790123457
6 75.623621399177
};
\addlegendentry{70}
\addplot [semithick, crimson2143940]
table {%
2 72.8584
3 75.6238
4 76.458
5 75.8674
6 75.4394
};
\addlegendentry{80}
\addplot [semithick, mediumpurple148103189]
table {%
2 72.7812345679012
3 75.4238271604938
4 76.258024691358
5 75.9279012345679
6 75.7541563786008
};
\addlegendentry{90}
\addplot [semithick, sienna1408675]
table {%
2 72.8383127572016
3 75.0724279835391
4 76.5577777777778
5 76.2285596707819
6 75.830987654321
};
\addlegendentry{100}
\end{axis}

\end{tikzpicture}}
        \caption{}
    \end{subfigure}
    \hfill
    \begin{subfigure}{0.3\linewidth}
        \centering
        \scalebox{0.6}{
\begin{tikzpicture}

\definecolor{darkgray176}{RGB}{176,176,176}
\definecolor{darkorange25512714}{RGB}{255,127,14}
\definecolor{lightgray204}{RGB}{204,204,204}
\definecolor{steelblue31119180}{RGB}{31,119,180}

\begin{axis}[
legend cell align={left},
legend style={
  fill opacity=0.8,
  draw opacity=1,
  text opacity=1,
  at={(0.03,0.97)},
  anchor=north west,
  draw=lightgray204
},
tick align=outside,
tick pos=left,
x grid style={darkgray176},
xlabel={Kernel Size},
xmin=1.1, xmax=20.9,
xtick style={color=black},
y grid style={darkgray176},
ylabel={Accuracy (\%)},
ymin=67.0320216049383, ymax=79.4787808641975,
ytick style={color=black}
]
\addlegendimage{empty legend}
\addlegendentry{\hspace{-.6cm}Evaluation setup:}
\addplot [draw=steelblue31119180, fill=steelblue31119180, forget plot, mark=*, only marks]
table{%
x  y
2  68.4614197530864
3  71.1234567901235
4  70.795524691358
5  71.5864197530864
6  71.1813271604938
7  71.054012345679
8  71.0308641975309
9  70.6527777777778
10 70.2438271604938
11 69.170524691358
12 69.3981481481482
13 69.0817901234568
};
\addplot [draw=darkorange25512714, fill=darkorange25512714, forget plot, mark=*, only marks]
table{%
x  y
5  72.658950617284
6  73.8472222222222
7  74.6188271604938
8  75.5802469135802
9  76.7067901234568
10 76.9768518518519
11 76.4475308641975
12 75.945987654321
13 77.4564814814815
15 78.2469135802469
16 77.5324074074074
17 77.0493827160494
18 75.9768518518518
19 76.2006172839506
20 76.3086419753087
};
\addplot [semithick, steelblue31119180]
table {%
2  68.4614197530864
3  71.1234567901235
4  70.795524691358
5  71.5864197530864
6  71.1813271604938
7  71.054012345679
8  71.0308641975309
9  70.6527777777778
10 70.2438271604938
11 69.170524691358
12 69.3981481481482
13 69.0817901234568
};
\addlegendentry{Cross-Subject}
\addplot [semithick, darkorange25512714]
table {%
5  72.658950617284
6  73.8472222222222
7  74.6188271604938
8  75.5802469135802
9  76.7067901234568
10 76.9768518518519
11 76.4475308641975
12 75.945987654321
13 77.4564814814815
15 78.2469135802469
16 77.5324074074074
17 77.0493827160494
18 75.9768518518518
19 76.2006172839506
20 76.3086419753087
};
\addlegendentry{Within-Subject}
\end{axis}

\end{tikzpicture}}
        \caption{}
    \end{subfigure}
    \caption{Impact of the variations of EEG-Simpleconv's hyperparameters of EEG-SimpleConv on the Cross-Subject classification accuracy on four datasets. (a) variation of the width $W$, (b) variation of the size of the convolutions kernels, (c) variation of the High-pass filter frequency, variation of the Depth $K$ on (d) BNCI, (e) Zhou, (g) Physionet (h) Cho.
    Impact of the variation of (f) the Depth $K$ and (i) the kernel size on Cross-Subject performances compared to Within-Subject on BNCI.}
    \label{fig:hyperparameters}
\end{figure*}

By considering these recommended parameter ranges, it will be faster to obtain optimal performance for new experiments. The results of this table were obtained with the experiment illustrated in Fig.~\ref{fig:hyperparameters}.

\section{Size and Inference Time further analysis}
\label{app:time_size}

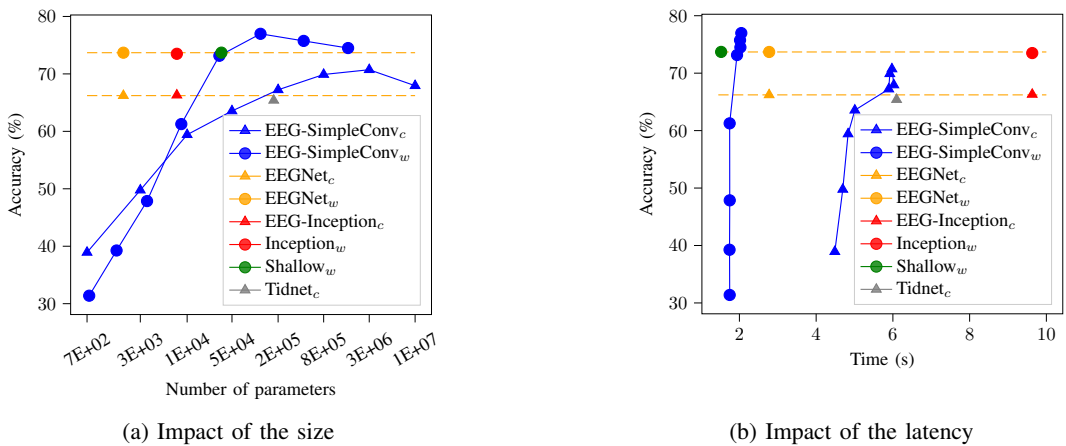
\begin{figure*}
    \centering
    \hspace*{\fill}
    \begin{subfigure}{0.45\linewidth}
        \centering
        \scalebox{0.7}{
\begin{tikzpicture}

\definecolor{darkgray176}{RGB}{176,176,176}
\definecolor{gray}{RGB}{128,128,128}
\definecolor{green}{RGB}{0,128,0}
\definecolor{lightgray204}{RGB}{204,204,204}
\definecolor{orange}{RGB}{255,165,0}

\begin{axis}[
legend cell align={left},
legend style={
  fill opacity=0.8,
  draw opacity=1,
  text opacity=1,
  at={(0.97,0.03)},
  anchor=south east,
  draw=lightgray204
},
tick align=outside,
tick pos=left,
x grid style={darkgray176},
xlabel={Number of parameters},
xmin=5.9894093804946, xmax=16.9179403663833,
xtick style={color=black},
xticklabel style={rotate=30},
xtick={6.48616078894409,8.09925056179696,9.51753091081601,10.8753265137097,12.2724781727268,13.6518521559848,15.0322009498632,16.4211889579339},
xticklabels={7E+02,3E+03,1E+04,5E+04,2E+05,8E+05,3E+06,1E+07},
y grid style={darkgray176},
ylabel={Accuracy (\%)},
ymin=28.1089248971193, ymax=80.2568158436214,
ytick style={color=black}
]
\path [draw=orange, semithick, dash pattern=on 5.55pt off 2.4pt]
(axis cs:6.48616078894409,73.69)
--(axis cs:16.4211889579339,73.69);

\path [draw=orange, semithick, dash pattern=on 5.55pt off 2.4pt]
(axis cs:6.48616078894409,66.22)
--(axis cs:16.4211889579339,66.22);

\addplot [semithick, blue, mark=triangle*, mark size=3, mark options={solid}]
table {%
6.48616078894409 38.9274691358025
8.09925056179696 49.7654320987654
9.51753091081601 59.4169238683128
10.8753265137097 63.5514403292181
12.2724781727268 67.2294238683128
13.6518521559848 69.897890946502
15.0322009498632 70.727366255144
16.4211889579339 67.943158436214
};
\addlegendentry{EEG-SimpleConv$_c$}
\addplot [semithick, blue, mark=*, mark size=3, mark options={solid}]
table {%
6.5510803350434 31.3883744855967
7.38025578842646 39.2587448559671
8.3030093814735 47.8621399176955
9.3410176956178 61.2623456790124
10.494048144321 73.1579218106996
11.7396147228309 76.977366255144
13.0473725541137 75.7427983539095
14.3919643589769 74.5
};
\addlegendentry{EEG-SimpleConv$_w$}
\addplot [semithick, orange, mark=triangle*, mark size=3, mark options={solid}]
table {%
7.58680353516258 66.22
};
\addlegendentry{EEGNet$_c$}
\addplot [semithick, orange, mark=*, mark size=3, mark options={solid}]
table {%
7.58680353516258 73.69
};
\addlegendentry{EEGNet$_w$}
\addplot [semithick, red, mark=triangle*, mark size=3, mark options={solid}]
table {%
9.20552881497896 66.3
};
\addlegendentry{EEG-Inception$_c$}
\addplot [semithick, red, mark=*, mark size=3, mark options={solid}]
table {%
9.20552881497896 73.5
};
\addlegendentry{Inception$_w$}
\addplot [semithick, green, mark=*, mark size=3, mark options={solid}]
table {%
10.5559168998175 73.7
};
\addlegendentry{Shallow$_w$}
\addplot [semithick, gray, mark=triangle*, mark size=3, mark options={solid}]
table {%
12.1390136172482 65.4
};
\addlegendentry{Tidnet$_c$}
\end{axis}

\end{tikzpicture}}
        \caption{Impact of the size}
    \end{subfigure}
    \begin{subfigure}{0.45\linewidth}
        \centering
        \raisebox{0.4cm}{\scalebox{0.7}{
\begin{tikzpicture}

\definecolor{darkgray176}{RGB}{176,176,176}
\definecolor{gray}{RGB}{128,128,128}
\definecolor{green}{RGB}{0,128,0}
\definecolor{lightgray204}{RGB}{204,204,204}
\definecolor{orange}{RGB}{255,165,0}

\begin{axis}[
legend cell align={left},
legend style={
  fill opacity=0.8,
  draw opacity=1,
  text opacity=1,
  at={(0.97,0.03)},
  anchor=south east,
  draw=lightgray204
},
tick align=outside,
tick pos=left,
x grid style={darkgray176},
xlabel={Time (s)},
xmin=1.0225, xmax=10.4275,
xtick style={color=black},
y grid style={darkgray176},
ylabel={Accuracy (\%)},
ymin=28.1089248971193, ymax=80.2568158436214,
ytick style={color=black}
]
\path [draw=orange, semithick, dash pattern=on 5.55pt off 2.4pt]
(axis cs:1.45,66.22)
--(axis cs:10,66.22);

\path [draw=orange, semithick, dash pattern=on 5.55pt off 2.4pt]
(axis cs:1.45,73.69)
--(axis cs:10,73.69);

\addplot [semithick, blue, mark=triangle*, mark size=3, mark options={solid}]
table {%
4.48626780509949 38.9274691358025
4.69391918182373 49.7654320987654
4.83233666419983 59.4169238683128
5.00778913497925 63.5514403292181
5.89227191925049 67.2294238683128
5.92001780700684 69.897890946502
5.97519111633301 70.727366255144
6.0247528553009  67.943158436214
};
\addlegendentry{EEG-SimpleConv$_c$}
\addplot [semithick, blue, mark=*, mark size=3, mark options={solid}]
table {%
1.74672245979309 31.3883744855967
1.74292063713074 39.2587448559671
1.74786496162415 47.8621399176955
1.74577045440674 61.2623456790124
1.93811058998108 73.1579218106996
2.04935193061829 76.977366255144
2.01630640029907 75.7427983539095
2.02584886550903 74.5
};
\addlegendentry{EEG-SimpleConv$_w$}
\addplot [semithick, orange, mark=triangle*, mark size=3, mark options={solid}]
table {%
2.775 66.22
};
\addlegendentry{EEGNet$_c$}
\addplot [semithick, orange, mark=*, mark size=3, mark options={solid}]
table {%
2.775 73.69
};
\addlegendentry{EEGNet$_w$}
\addplot [semithick, red, mark=triangle*, mark size=3, mark options={solid}]
table {%
9.632 66.3
};
\addlegendentry{EEG-Inception$_c$}
\addplot [semithick, red, mark=*, mark size=3, mark options={solid}]
table {%
9.632 73.5
};
\addlegendentry{Inception$_w$}
\addplot [semithick, green, mark=*, mark size=3, mark options={solid}]
table {%
1.524 73.7
};
\addlegendentry{Shallow$_w$}
\addplot [semithick, gray, mark=triangle*, mark size=3, mark options={solid}]
table {%
6.096 65.4
};
\addlegendentry{Tidnet$_c$}
\end{axis}

\end{tikzpicture}}}
        \caption{Impact of the latency}
    \end{subfigure}
    \hspace*{\fill}
    \caption{Impact of the size and the latency on the classification performance of EEG-SimpleConv. x-axis of (a) is in log-scale.}
    \label{fig:time_size}
\end{figure*}

In Fig.~\ref{fig:time_size}, we examine the effect of the size and inference time of EEG-SimpleConv on its classification performance, while varying the number of feature maps $W$. We observe that the model inference time evolves slowly while increasing the number of feature maps.

\end{document}